\documentclass[fleqn,usenatbib]{mnras}

\usepackage{amssymb}

\usepackage[T1]{fontenc}

\DeclareRobustCommand{\VAN}[3]{#2}
\let\VANthebibliography\thebibliography
\def\thebibliography{\DeclareRobustCommand{\VAN}[3]{##3}\VANthebibliography}

\usepackage{graphicx}	
\usepackage{amsmath}	
\usepackage{amssymb}	

\usepackage{newtxtext,newtxmath}

\title[Image quality and merger identification]{The effect of image quality on galaxy merger identification with deep learning}

\author[R. W. Bickley et al.]{Robert W. Bickley,$^{1}$\thanks{E-mail: rw\_bickley@proton.me}
Scott Wilkinson,$^{1}$
Leonardo Ferreira,$^{1}$
\newauthor
Sara L. Ellison,$^{1}$
Connor Bottrell,$^{2}$
Debarpita Jyoti$^{3}$
\\
$^{1}$Department of Physics and Astronomy, University of Victoria, Victoria, British Columbia V8P 1A1, Canada\\
$^{2}$International Centre for Radio Astronomy Research, University of Western Australia, 35 Stirling Hwy, Crawley, WA 6009, Australia\\
$^{3}$ Department of Physics, Indian Institute of Technology, Kharagpur, India\\
}

\date{Accepted XXX. Received YYY; in original form ZZZ}

\pubyear{2024}

\begin{document}
\label{firstpage}
\pagerange{\pageref{firstpage}--\pageref{lastpage}}
\maketitle

\begin{abstract}
Studies have shown that the morphologies of galaxies are substantially transformed following coalescence after a merger, but post-mergers are notoriously difficult to identify, especially in imaging that is shallow or low-resolution. We train convolutional neural networks (CNNs) to identify simulated post-merger galaxies in a range of image qualities, modelled after five real surveys: the Sloan Digital Sky Survey (SDSS), the Dark Energy Camera Legacy Survey (DECaLS), the Canada-France Imaging Survey (CFIS), the Hyper Suprime-Cam Subaru Strategic Program (HSC-SSP), and the Legacy Survey of Space and Time (LSST). Holding constant all variables other than imaging quality, we present the performance of the CNNs on reserved test set data for each image quality. The success of CNNs on a given dataset is found to be sensitive to both imaging depth and resolution. We find that post-merger recovery generally increases with depth, but that limiting 5$\sigma$ point-source depths in excess of $\sim25$ mag, similar to what is achieved in CFIS, are only marginally beneficial. Finally, we present the results of a cross-survey inference experiment, and find that CNNs trained on a given image quality can sometimes be applied to different imaging data to good effect. The work presented here therefore represents a useful reference for the application of CNNs for merger searches in both current and future imaging surveys.
\end{abstract}

\begin{keywords}
Galaxies: Evolution -- Galaxies: Interactions -- Galaxies: Peculiar -- Methods: Statistical -- Techniques: Image Processing
\end{keywords}



\section{Introduction}
\label{Introduction}

After merging galaxies coalesce into a single post-merger remnant, simulations predict that they are likely to experience an epoch of rapid change. When the merger is gas-rich, morphological disturbances (e.g., as studied in \citealp{1972ApJ...178..623T,1993ASPC...48..615B,2008MNRAS.391.1137L,2017MNRAS.465.1106B,2024MNRAS.528.2326S}) in the post-merger are expected to drive gas towards the centre. The elevated gas densities in turn give rise to elevated central star formation rates (SFRs; \citealp{Mihos1996GasdynamicsMergers,2008AN....329..952D,10.1093/mnras/stt017,2014A&A...566A..97J,2015MNRAS.448.1107M,2019MNRAS.490.2139R,2023MNRAS.522.5107B}) and supermassive black hole (SMBH) accretion rates (\citealp{2005MNRAS.361..776S,2017ApJ...845..128P,2022MNRAS.517.4752S,2023MNRAS.519.4966B,2024MNRAS.528.5864B}). SMBH accretion is an energetic phenomenon, and energy injection by the SMBH, often referred to as active galactic nucleus (AGN) feedback, can lead to regulation, suppression, and even truncation of star formation in post-merger galaxies (\citealp{2009ApJ...690..802J,2014MNRAS.437.1456B,2014MNRAS.442..440C,2022MNRAS.515.1430D,2022MNRAS.513...27Z,2023MNRAS.519.2119Q}).

Many of these theoretical predictions have been borne out by observations. Multiple observational studies have found evidence of elevated star formation rates (\citealp{2013MNRAS.435.3627E,2016RAA....16..113G,2019MNRAS.482L..55T,2022MNRAS.516.1462T,2019ApJ...881..119P,2020ApJ...902...77O,2023PASJ...75..986T,2024MNRAS.527.2037R}), SMBH activity (\citealp{2013MNRAS.435.3627E,2019MNRAS.487.2491E,2014MNRAS.441.1297S,2017MNRAS.464.3882W,2020A&A...637A..94G,2023ApJ...944..168L,2024ApJ...963...53C}), and eventual truncation of star formation (\citealp{2022MNRAS.517L..92E,2022ApJ...941...93O,2023MNRAS.523..720L}) in post-merger galaxies.

In observations, the presence and strength of the statistical connection between coalescence after a merger event and the subsequent physical metamorphosis is highly sensitive to the method by which the merger sample is selected. Indeed, a number of studies do not find a statistically significant link between elevated star formation, AGN detection, and mergers (e.g., \citealp{2011ApJ...726...57C,2012ApJ...744..148K,2012MNRAS.425L..61S,2019A&A...626A..49P,2019MNRAS.483.2441V,2020MNRAS.499.2380S}).

A variety of methods have been used to identify post-mergers in observations, including manual inspection (e.g., \citealp{2010ApJS..186..427N,2013MNRAS.435.3627E,2023ApJ...944..168L}), non-parametric morphological statistics (e.g., \citealp{2003yCat..21470001C,2004AJ....128..163L,2016MNRAS.456.3032P,2019MNRAS.483.4140R}), machine learning-based combinations of statistics (e.g., \citealp{2019ApJ...872...76N,2024MNRAS.528.5558W}), and deep learning methods (\citealp{2018MNRAS.479..415A,2019MNRAS.483.2968W,2019A&A...626A..49P,2020ApJ...895..115F,2020A&A...644A..87W}). Each approach has benefits and drawbacks, with each method generally serving to identify post-merger samples that are degrees of pure (i.e., containing relatively few non-post-merger interlopers), complete (containing a large proportion of the actual post-mergers in the parent sample being studied), and representative (containing galaxies that represent the typical demographical distribution of true post-mergers). Purity and representativeness are essential for the recovery of quantitatively accurate results about post-mergers, and completeness benefits statistical power.

\begin{table*}
\small
\centering
\begin{tabular}{ |p{4cm}|p{5cm}|p{1cm}|p{3.5cm}|p{1.75cm}| } 
\hline
Publication & \ Training set & \ Survey & \ Model & \ Completeness \\
\hline
\hline
\text{\citet{2019A&A...626A..49P}} & EAGLE pre- and post-mergers, matched controls & SDSS & \text{\citet{2015MNRAS.450.1441D}}& 0.65 \\
\hline
\text{\citet{2020A&A...644A..87W}} & TNG 100-1 pre- and post-mergers, matched controls & KiDS & \text{\citet{2014arXiv1409.1556S}} & 0.72 \\
\hline
\text{\citet{2022MNRAS.511..100B}} & TNG 100-1 post-mergers, matched controls & None & ResNet38-V2 & 0.93 \\
\hline
\text{\citet{2022A&A...661A..52P}} & Galaxy Zoo \text{\citep{2008MNRAS.389.1179L}} responses & KiDS & Custom CNN+ANN & 0.86\\
\hline
\text{\citet{2022ApJ...931...34F}} & TNG 100-1 post-mergers, matched star-forming galaxies & Hubble & Custom CNN & 0.80\\
\hline
\text{\citet{2022MNRAS.509.3966W}} & Galaxy zoo responses, multiple categories & DECaLS & Zoobot, \text{\citet{2019arXiv190511946T}} & 0.88\\
\hline
\text{\citet{2023MNRAS.521.3861D}} & NewHorizon galaxies with or without visible tidal features & HSC & Custom CNN & 0.85\\
\hline
\text{\citet{2023A&A...679A.142O}} & TNG 50 pre- and post-mergers, matched controls & HSC & \text{\citet{2022MNRAS.509.3966W}} & 0.76\\
\hline
\text{\citet{2024MNRAS.528.6915A}} & TNG 100-1 post-mergers, matched controls & SDSS & Custom CNN & 0.81\\
\hline
\text{\citet{2024A&A...687A..24M}} & TNG 100-1 pre- and post-mergers, non-mergers & HSC & \text{\citet{2022MNRAS.509.3966W}} & 0.78\\
\hline
\citet{2024arXiv240718396F} & TNG 100-1 pre- and post-mergers, non-mergers & CFIS & Multi-model ensemble & 0.84\\
\hline
\hline
\end{tabular}
\caption{The performance reported by several other works for simulation-trained deep learning identification of galaxy mergers. Efforts in the literature use a wide range of training sets; some use real, pre-labeled galaxy images, while others use a simulation-based approach. The data also come from multiple surveys, or use varying implementations of observational realism. The models used for post-merger classification also vary, but custom CNNs with architectures like the one used in this work are popular.}
\label{comparison_table}
\end{table*}

Deep learning methods like convolutional neural networks (CNNs) and Vision Transformers have been proven useful by a number of standards for post-merger identification (e.g., in \citealp{2019MNRAS.490.5390B,2020A&A...644A..87W,2021MNRAS.504..372B,2024MNRAS.528.6915A,2024arXiv240718396F}). They can quickly make predictions of merger status for large image samples, and appear to be capable of identifying complete post-merger samples. A number of efforts to use machine learning for merger classification are summarized in Table~\ref{comparison_table}. Even though they can be very accurate, single deep learning models (e.g., the approach used in \citealp{2021MNRAS.504..372B}; \citealp{2022MNRAS.511..100B}) are severely disadvantaged by the naturally low incidence rate of true post-mergers in the low-$z$ Universe. Generally, additional filtration of the predicted post-merger sample by visual classifiers (e.g., in \citealp{2022MNRAS.514.3294B,2024arXiv240117277B}) or using a multi-model classification framework (e.g., \citealp{2024arXiv240718396F,2024arXiv240718238L,2024A&A...687A..45P}) can be used to improve on the purity of lone classifiers. Multi-model frameworks are nevertheless sensitive to the performance of their constituent individual classifiers. In this work, we therefore focus on the performance of single CNNs for post-merger classification.

Several studies have adopted a simulation-based approach to post-merger searches by training machine vision tools on images of galaxies from cosmological hydrodynamical box simulations (e.g., \citealp{2019A&A...626A..49P,2020A&A...644A..87W,2021MNRAS.504..372B,2022MNRAS.511..100B,2022ApJ...931...34F,2023MNRAS.521.3861D,2023A&A...679A.142O,2024MNRAS.528.6915A,2024A&A...687A..24M}). In this work, we train models using galaxy stellar morphologies from the 100-1 run of the IllustrisTNG simulation suite (\citealp{2018MNRAS.480.5113M}; \citealp{2018MNRAS.477.1206N}; \citealp{2018MNRAS.475..624N}; \citealp{2018MNRAS.475..648P}; \citealp{2018MNRAS.475..676S}; \citealp{2019ComAC...6....2N}).

The completeness and representativeness of a post-merger sample identified by a simulation-trained CNN are limited by the degree of realism of galaxies from the simulation, and the quality of observational realism created when mock observations of simulated galaxies are performed. The limiting factors are already explored in some depth in the literature. For example, \citet{2019MNRAS.483.4140R} , who demonstrate $1\sigma$ statistical agreement in the non-parametric morphological measurements taken from galaxies in the low-$z$ Universe and galaxies from IllustrisTNG. \citet{2024MNRAS.528.7411E}, meanwhile, find that there is a 70 per cent overlap in the parameter space uncovered via contrastive learning between real low-$z$ galaxies and IllustrisTNG galaxies. \citet{2024arXiv240718396F} also demonstrate that Uniform Manifold Approximation and Projection (UMAP) representations of the parameter spaces occupied by CFIS and IllustrisTNG galaxies with CFIS realism are in good qualitative agreement. \citet{2019MNRAS.490.5390B} verified that observational realism is important in training CNNs that will be eventually applied to real observational datasets. While the realistic appearances of galaxies in IllustrisTNG and in the training set place limitations on the maximum potential of our approach, they are not so much a hinderance as to prevent the successful identification of mergers in the low-$z$ Universe (see \citealp{2022MNRAS.514.3294B}).

Image quality has previously been discussed as a factor in the success of morphological classification (e.g., \citealp{2004AJ....128..163L,2008ApJS..179..319L}), and efforts have also been made in the literature to characterize the benefits of current and next-generation imaging surveys. For example, \citet{2022MNRAS.513.1459M} investigate the potential observability of low-surface brightness features in the Legacy Survey of Space and Time (LSST; \citealp{2019ApJ...873..111I}) taken at the Vera Rubin observatory, and \citet{2023MNRAS.521.3861D} perform a similar assessment using CNNs to identify tidal features in Hyper Suprime-Cam (\citealp{10.1093/pasj/psab122}) imaging. The most direct literature precursor to this study, \citet{2024MNRAS.528.5558W}, contains a systematic analysis of merger observability on a grid of depth and resolution, using individual and machine-combined non-parametric morphological statistics. Better (i.e., deeper, higher resolution) imaging data is generally considered to be beneficial, but imaging quality must sometimes be sacrificed for quantity (i.e., extent of on-sky coverage) in pursuit of large merger samples and scientific results with good statistical significance. \citet{2022MNRAS.515.3406M} conduct a similar investigation of the role of field of view in successful merger identification using non-parametric morphological statistics. Meanwhile, \citet{2018MNRAS.473.2701D} explore the specific connection between redshift and morphology in several imaging surveys, identifying suitable $z$ ranges for robust morphological characterization. As large, high-quality imaging data from multiple observatories are more often becoming widely available, survey depth and resolution are becoming factors that astronomers can choose when designing experiments. Still greater degrees of control will be available in coming years, with data from LSST becoming available. In this work, we will explore the extent to which the success of low-$z$ post-merger searches is limited by imaging, using five real surveys -- SDSS, DECaLS, CFIS, the Hyper Suprime-Cam Subaru Strategic Program Wide field (HSC-W), and the projected 10-year depth of LSST (hereafter referred to simply as LSST) -- to sample the technical parameter space broadly referred to as ``imaging quality''.

\section{Data and methods}
\label{Data and methods:realism}

Several methodological improvements and insights have been developed since our original work presented in \citet{2021MNRAS.504..372B}, so our approach to this comparative study is described here in detail, even though the experiments are philosophically similar (involving training CNNs on mock images of post-merger and control galaxies from IllustrisTNG100-1). In this Section, we will explain how post-merger and non-post-merger control galaxies are selected from the simulation, how mock observations of the galaxies are performed, and how the CNN is trained. We will also detail how predictions of merger status are made.

\subsection{IllustrisTNG 100-1 galaxy selection}
\label{IllustrisTNG 100-1 galaxy selection}

We use matched samples of post-merger galaxies and control galaxies to explore the potential of CNNs as a function of image quality. The selection for both classes is updated from that used in \citet{2021MNRAS.504..372B}, and is designed to select a somewhat smaller number of galaxies that are more unambiguously post-mergers or non-post-mergers.

\subsubsection{Post-mergers}
\label{Post-mergers:realism}

The post-mergers and control galaxy selection for this work is based on that of \citet{2024arXiv240718396F}. \citet{2024arXiv240718396F} themselves use the \citet{2024MNRAS.528.5864B} approach of searching the IllustrisTNG 100-1 simulation merger trees to determine merger status. As in \citet{2021MNRAS.504..372B}, \citet{2024arXiv240718396F} use galaxy metadata from IllustrisTNG to select galaxies from snapshots between simulation redshifts $0<z<1$. \citet{2024arXiv240718396F} argue that galaxy morphologies outside of this redshift range are expected to be statistically and qualitatively different from those found at low-$z$.

Simulation quality control cuts are also enforced; \textsc{SubhaloFlag=True} ensures that galaxies have dark matter and total mass properties consistent with cosmological origin, and log$(\mathrm{M_{\star}/M_{\odot}})>10$ ensures that each galaxy contains at least $\sim7000$ star particles. Departing from the selection used in \citet{2021MNRAS.504..372B}, there is also an upper mass cut at log$(\mathrm{M_{\star}/M_{\odot}})<11$. Since a disproportionate number of galaxies with high stellar masses are experiencing a merger, it is difficult to match a representative sample of non-merging control galaxies to them. From this subset \citet{2024arXiv240718396F} select a general merger sample with stellar mass ratios $\mu>$1:10, and identify a broad post-merger class including all galaxies that have coalesced after a merger that took place in the last 1.7 Gyr. The 1.7 Gyr time window for merger selection is designed to capture the vast majority of galaxies that might exhibit post-merger-like morphologies, since mergers may be identifiable up to 2 Gyr after coalescence (see \citealp{2008MNRAS.391.1137L}.) \citet{2024arXiv240718396F} again follow \citet{2024MNRAS.528.5864B} and use an updated measurement of $\mu$, which considers the mass ratio between interacting companion galaxies at a time when they were at least 50 kpc apart. In comparing the masses of interacting galaxies while they are still well separated, one avoids the effects of ``numerical stripping'', a simulation problem in which galaxy masses can be under-estimated when particles are wrongly associated with a nearby neighbour. For this work, we take the subset of immediate post-mergers for which coalescence took place in the simulation snapshot prior to imaging (i.e., more than one progenitor subhalo from the previous snapshot are associated with a new subhalo in the present snapshot). Algorithmically-identified post-mergers in IllustrisTNG have a range of characteristics, with some examples still hosting more than one stellar nucleus, and others having the appearance of a more complete coalescence. Visual inspection of the post-merger sample indicates that cases with multiple nuclei are separated by $<10$ kpc, and share a common stellar envelope (conveniently, a common criterion for visual identification of post-mergers). The final sample of post-mergers for this work contains 1627 galaxies. The main differences between the post-merger sample used in this paper and that used in \citet{2021MNRAS.504..372B} are the upper limit on $\mathrm{M_{\star}}$, the updated $\mu$ estimate for the simulation, and the enforcement of the \textsc{SubhaloFlag}.

\subsubsection{Controls}
\label{Controls:realism}

An equal number (1627) of control galaxies are matched to the post-mergers described in Section~\ref{Post-mergers:realism}. The control pool includes two types of non-merger galaxies: ``true'' and ``potential''. ``True'' non-mergers have not experienced a merger with a mass ratio of $>$1:10 in the last 1.7 Gyr ($T_{\text{post-merger}}>1.7$), and do not experience a merger for at least 1.7 Gyr after the time of imaging ($T_{\text{until-merger}}>1.7$). ``Potential'' non-mergers have not experienced a merger with a mass ratio of $>$1:10 in the last 1.7 Gyr, and are unlikely to experience a merger event in the next few Gyr based on a nearest neighbour separation criterion ($r_{1}>50$ kpc). The mass ratio criterion for controls means that more minor mergers are included in the control sample. The potential non-mergers are distinct from true non-mergers because the simulation runtime came to an end before their merger trees can be fully inspected for potential future merger events. Both types of non-mergers are included in the control pool from which the matched controls in this work are drawn.

\begin{figure*}
\includegraphics[width=\textwidth]{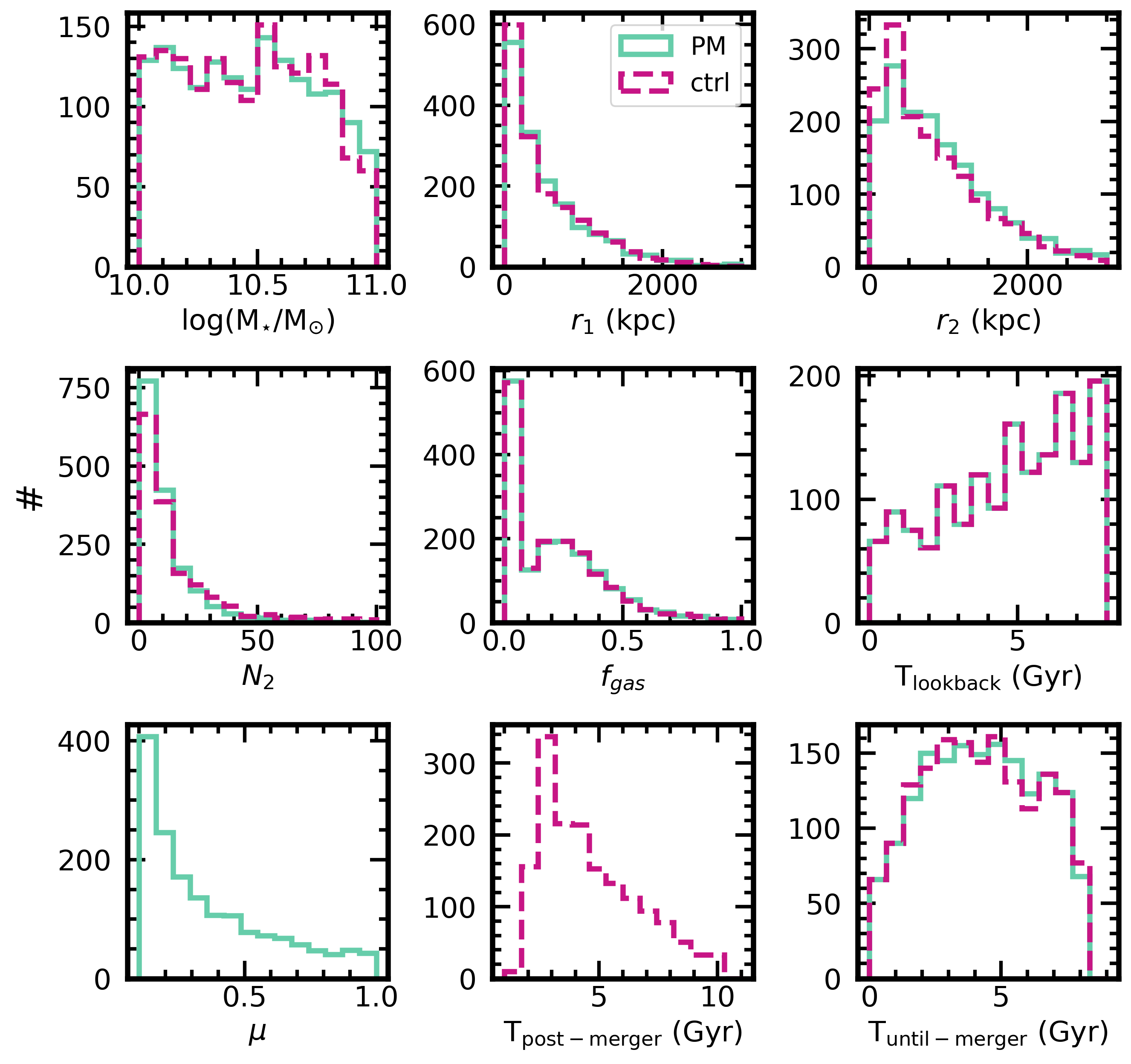}
\caption{The stellar mass, lookback time, gas fraction ($f_{\text{gas}}$), mass ratio $\mu$, and environment ($r_{1}$, $r_{2}$, $N_{2}$) statistics for the post-mergers (teal) and non-merger controls (magenta) used to train the CNNs in this work. We also include $T_{\text{post-merger}}$ for the controls (but not for the post-mergers because they all have $T_{\text{post-merger}}=0$) and $T_{\text{until-merger}}$ for both classes.}
\label{fig:multi-hist-realism}
\end{figure*}

Individual control galaxies are matched to the post-mergers on three parameters: $\mathrm{M_{\star}}$, $z$, and gas fraction $f_{\text{gas}}$, where $f_{\text{gas}}$ is a unitless quantity defined as the gas mass divided by the sum of the stellar and gas masses. The $\mathrm{M_{\star}}$ matching tolerance is 0.05 dex, and the $f_{\text{gas}}$ tolerance is 0.05. Control galaxies must also be taken from the same simulation snapshot (i.e., with identical lookback time, $T_{\text{lookback}}$) as their post-merger counterparts. Gas content and the availability of gas for forming new stars is known to affect the observed morphologies of post-mergers; see \citet{2024MNRAS.528.5558W}, \citet{2010MNRAS.404..575L}, and \citet{2006ApJ...652..270B}. The nearest neighbour in the $\mathrm{M_{\star}}-f_{\text{gas}}$ plane is taken as the best control for each merger. Controls are matched without replacement so that the final samples of mergers and controls contain the same number of unique galaxies. Figure~\ref{fig:multi-hist-realism} shows that the merger identification and control matching procedures have been effective in identifying statistically comparable galaxy samples whose primary difference is merger status. Figure~\ref{fig:multi-hist-realism} also shows the sample statistics for a number of parameters not used for control matching, including second nearest neighbour distance ($r_{2}$), the number of neighbour galaxies within 2 Mpc ($N_{2}$).

\subsection{Mock observations}
\label{Mock observations}

The mock observation approach used in this work is similar to that in \citet{2021MNRAS.504..372B}, with small modifications. First, redshifts are no longer chosen at random from a real population of galaxy redshifts. Instead, we identify five redshift bins between $0<z<0.3$ that contain equal numbers of galaxies in the seventh data release (DR7) of the Sloan Digital Sky Survey (SDSS; \citealp{2000AJ....120.1579Y}). The redshift values in the centres of each bin are $z={0.036,0.101,0.150,0.198,0.256}$. At each of the five redshifts, we perform one mock observation with a fixed physical diameter of 100 kpc for each of four camera angles at the vertices of an equilateral tetrahedron with the galaxy at its centre. We therefore make a total of 20 (five redshifts times four camera angles) mock observations of each galaxy.

Since real skies are not available from all of the surveys we plan to study (10-year imaging co-adds from LSST are not yet available at the time of writing), we use synthetic image backgrounds based on the reported observational parameters for each survey. We repeat the mock observation procedure for each of the five surveys, so that the complete image dataset used in this work includes 20 images for each of 1627 post-mergers and 1627 controls (a total of 65080 images) for each of five surveys. The observational parameters for each of the surveys are outlined in Table~\ref{survey-params-table}, including the charge-coupled device (CCD) pixel scale (on-sky angle subtended by a single pixel), the PSF (expected or known instrumental PSF for the survey, dominated by the atmosphere), $\sigma_{\text{sky}}$ (the standard deviation of the Gaussian sky noise in magnitudes per square arcsecond used to generate the mock observation), and the $5\sigma$ point-source depth (a convenient quantity reported as an at-a-glance metric of the limiting depth for many surveys). The essential statistics used to determine $\sigma_{\text{sky}}$ for each survey were found at the following sources: SDSS\footnote{https://classic.sdss.org/dr7/}, DECaLS (\citealp{2019AJ....157..168D}), CFIS DR5\footnote{https://www.cadc-ccda.hia-iha.nrc-cnrc.gc.ca/ \newline en/community/unions/MegaPipe\_CFIS\_DR3.html}, HSC-W (\citealp{10.1093/pasj/psab122}), and the LSST 10-year co-adds (\citealp{2019ApJ...873..111I}).

\begin{table}
\begin{center}
\begin{tabular}{ |c|c|c|c|c| } 
\hline
Survey & \ CCD pixel scale & \ PSF & \ $\sigma_{\text{sky}} $ & \ $5\sigma$ PSD \\
\hline
(units) & \ (arcsec/pix) & \ (arcsec) & \ (mag/$\mathrm{arcsec}^{2}$) & \ (mag) \\
\hline
\hline
SDSS & 0.396 & 1.4 & 24.33 & 22.7 \\
\hline
DECaLS & 0.262 & 1.18 & 24.97 & 23.54 \\
\hline
CFIS DR5 & 0.186 & 0.69 & 25.6 & 25.0 \\
\hline
HSC-W & 0.17 & 0.75 & 26.95 & 26.5 \\
\hline
LSST 10y & 0.2 & 0.7 & 28.12 & 27.5 \\
\hline
\hline
\end{tabular}
\end{center}
\caption{Essential parameters describing the image quality of $r$-band data from each of the surveys included in this comparison. The $5\sigma$ point-source depths and PSFs are plotted against one another later in Figure~\ref{fig:depth_res_grid}.}
\label{survey-params-table}
\end{table}

A new approach for generating synthetic sky backgrounds was developed for this effort. First, the light from the synthetic galaxy is re-binned to the required CCD pixel scale and blurred with an artificial Gaussian PSF with the same angular size as reported in the literature (shown in Table~\ref{survey-params-table}). The implementation of PSF blurring is the same for each of the five surveys. Next, the parameter $\sigma_{\text{sky}}$ is needed to specify the amplitude of the noise that will be added to approximate the sky. Computing $\sigma_{\text{sky}}$ analytically without access to imaging is non-trivial, since $\sigma_{\text{sky}}$ is sensitive to the CCD pixel scale, the PSF, the survey's true sensitivity (approximated here by the $5\sigma$ point source depth), and the aperture within which the signal (e.g., from a galaxy image or a point source) is compared to the noise. In order to determine $\sigma_{\text{sky}}$ in a homogeneous way for each of the five surveys, we used the following procedure:

\begin{itemize}
\item Create a point-source with a brightness equal to the reported limiting $5\sigma$ point-source depth.
\medskip
\item Convolve the source with the PSF of the relevant survey.
\medskip
\item Add the source to a test image containing an arbitrarily high level of Gaussian noise at the pixel scale of the survey.
\medskip
\item Using \textsc{SourceExtractor} (\citealp{1996A&AS..117..393B}) as implemented in Python by \citet{2016JOSS....1...58B}, measure the flux and flux error within some circular aperture.
\medskip
\item Repeat all previous steps with incrementally less noise in steps of 0.025 mag arcsec$^{-2}$ until the flux divided by the flux error is at least 5 (i.e., $5\sigma$).
\medskip
\item Repeat all previous steps 500 times and take median to ensure there is no dependence on noise.
\medskip
\item If the aperture used to estimate the $5\sigma$ point source depth is not given, repeat all previous steps with varying circular apertures until the flux (at the point of a $5\sigma$ detection) is equal to the reported $5\sigma$ limiting point source depth of the survey in magnitudes.
\medskip
\item Report the standard deviation used to generate the final sky background as $\sigma_{\text{sky}}$.
\end{itemize}

Since the modified version of \textsc{RealSim} \citep{2019MNRAS.490.5390B} used to generate mock CFIS images in \citet{2021MNRAS.504..372B} accepts a false sky $\sigma$, CCD pixel scale, and PSF as input, the availability of $\sigma_{\text{sky}}$ means that image quality is now fully parameterized. Other than the modifications to the redshift selection and false sky generation methods, the mock observation and image normalization procedures are the same as outlined in Section 2.2 of \citet{2021MNRAS.504..372B}, in which surface brightness maps are computed from the stellar mass surface density of each galaxy and the $r$-band absolute magnitudes for IllustrisTNG galaxies tabulated in \citet{2018MNRAS.475..624N}.

\begin{figure*}
\includegraphics[width=\textwidth]{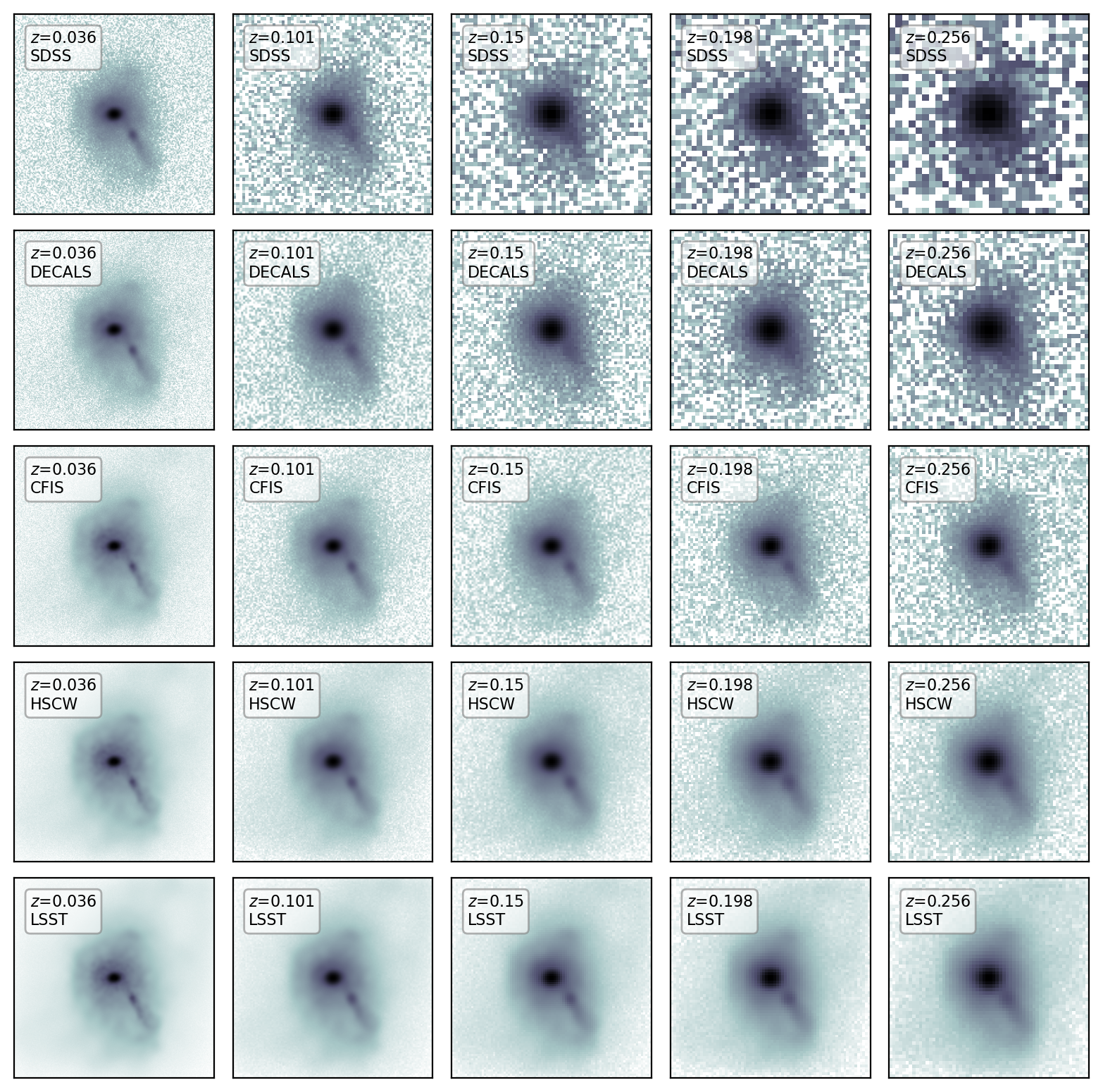}
\caption{A mosaic showing all of the survey and redshift realizations for one camera angle for one galaxy from the post-merger sample. Images are 100 kpc on a side. Redshift increases from left to right, and particularly in the shallower surveys studied in this work (e.g., SDSS, DECaLS) the decrease in visibility of the merger-induce tidal tail with increasing mock observation $z$ can be noted. The differences between rows highlight the importance of PSF, CCD scale resolution, and limiting depth in preserving the morphological signatures of a recent merger event. The galaxy images are shown with $1\sigma$ log-normalized scaling to better highlight the low-surface brightness features.}
\label{fig:all_ims_merger}
\end{figure*}

\begin{figure*}
\includegraphics[width=\textwidth]{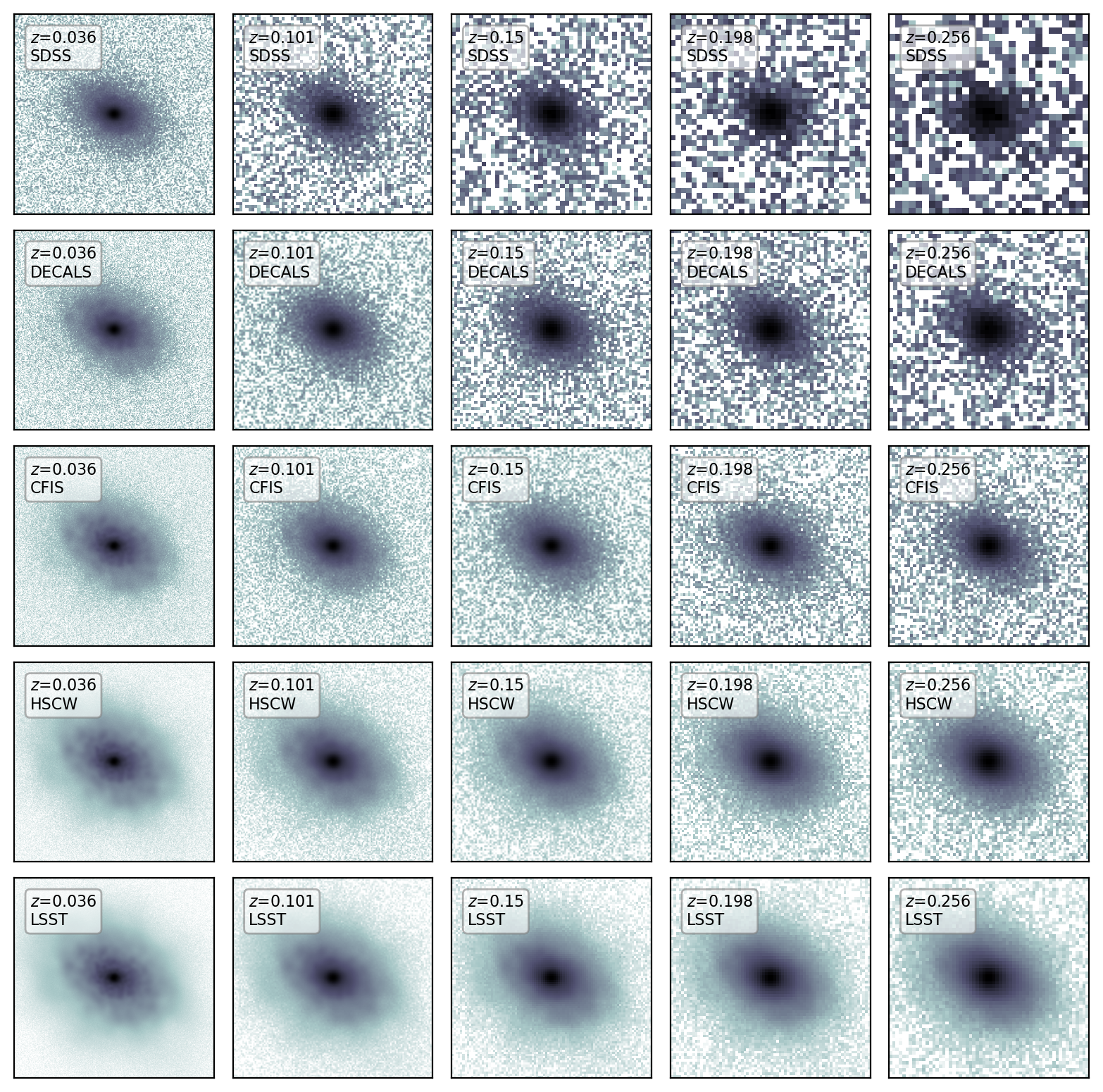}
\caption{The same as Figure~\ref{fig:all_ims_merger}, but for one galaxy from the control sample. Even for an undisturbed galaxy, substructure detail becomes increasingly visible for surveys with deeper limiting magnitudes or finer resolution.}
\label{fig:all_ims_ctrl}
\end{figure*}

Figures~\ref{fig:all_ims_merger} and~\ref{fig:all_ims_ctrl} show a useful cross-section of the data used to train and evaluate the CNNs in this work. In each row, both Figures show the five $z$ realizations created for a single camera angle of one post-merger (for Figure~\ref{fig:all_ims_merger}) and one control (for Figure~\ref{fig:all_ims_ctrl}) in a single survey. Subsequent rows show the corresponding images from each of the five surveys considered. In SDSS, the visibility of the prominent tidal tail in the post-merger galaxy is hidden in the image noise by $z=0.2$, and the post-merger and control look indistinguishable from one another in SDSS in our highest-$z$ realization. The other surveys retain the low-surface brightness features associated with the merger across the redshift range, but the deepest images from the LSST camera (fourth row in each mosaic) allow for obvious distinction between the two classes even at the highest $z$ studied here.

\subsection{CNN architecture and training strategy}
\label{CNN architecture and training strategy}

We train CNNs on image data prepared according to the realism parameters of each survey. Images for training are partitioned into training, test, and validation sets, which represent 80, 10, and 10 per cent of the data, respectively. All 20 images associated with each post-merger galaxy, and all 20 images associated with the matched control for a given post-merger galaxy, only ever appear within one partition. The aim of this strategy is to prevent any amount of cross-contamination between the partitions, which could give rise to artificially high performance metrics. Moreover, the images corresponding to each galaxy appear within the same partition for all five surveys. In this way, each of the five CNNs is trained on nearly identical data, with the only difference being image quality. Controlling for as many nuisance parameters as possible (e.g., information leaking between data partitions, small fluctuations in performances due to differently-shuffled data) maximizes the significance of the comparison between the CNNs conducted later in Section~\ref{Results:realism}.

The CNN architecture used in this work is identical to that used in \citet{2021MNRAS.504..372B} except for the dimensions of the input layer. Since a $138\times138$-pixel input image was chosen in \citealp{2021MNRAS.504..372B} based on the $z$ demographics of galaxies in CFIS, it is not strictly fair to enforce an input image size of $138\times138$ for all five surveys, whose selection functions will include galaxies with different $z$ distributions than CFIS. Images in this work are resized instead to $128\times128$ pixels using the anti-aliasing resize algorithm from \textsc{skimage}\footnote{scikit-image.org} (\citealp{scikit-image}), and the final images are large enough to capture the essential detail in the galaxy images. An image size of $128\times128$ pixels is scientifically arbitrary, but is a popular choice in fixed-input machine vision problems where max pooling operations are used. The resizing operation has no bearing on the pixel-wise sensitivity of the images, since the \textsc{skimage} resize algorithm stretches or shrinks images, rather than re-binning them. Before training, the images are normalized in linear fashion so that their faintest pixel value is fixed at zero and the brightest is fixed at one. The same on-the-fly augmentation algorithm as in \citet{2021MNRAS.504..372B}, which applies translational, rotational, and shear transformations of at most 10 per cent during training, is used here as well. All models are allowed to train for an arbitrarily long time as long as their performance is improving -- we monitor the value of the loss function (binary cross-entropy, which characterizes cross contamination between the post-merger and control classes), and if the loss does not improve for 50 training epochs\footnote{One epoch represents a complete study of the training dataset, followed by a performance check on the validation set.}, training is terminated and the model weights from the epoch with the best validation performance are restored. Training is accelerated using multiple graphics processing units (GPUs) and the ``mirrored strategy'' approach currently being implemented in \textsc{tensorflow}\footnote{https://www.tensorflow.org/api\_docs/python/tf/distribute/MirroredStrategy}, in which the responsibilities of training are parallelized across multiple GPUs. Importantly, other than a factor decrease in training time proportional to the number of cores used, the models behave the same whether or not the mirrored strategy is used.

We tested two other architectures on the deepest and shallowest image datasets (LSST and SDSS, respectively) in order to assess whether the performance achieved in this work was limited by the data (this is the goal, as we are interested in the influence of data on the accuracy of the classifiers) or by the model. Implementations of AlexNet (\citealp{NIPS2012_c399862d}) and EfficientNetB0\footnote{keras.io/api/applications/efficientnet/} architecture performed similarly (within $\sim2$ per cent in accuracy for both post-mergers and controls) to what is reported in Section~\ref{Results:realism}. The models' similar global performance suggests that the results of this study are data-limited, rather than model-limited.

After training, each of the five CNNs are evaluated on the 10 per cent of data reserved for testing. Since our training classes contain equal numbers of images, we consider CNN predictions of $p(x)>0.5$ to indicate a post-merger classification. Finally, we perform an exhaustive cross-survey inference effort in which each model is evaluated on the other four training datasets (e.g., the SDSS model is evaluated on the SDSS, DECaLS, CFIS, HSC-W, and LSST test datasets). The model classifications taken from each of 25 inference sets (five models times five test sets) constitute the new material used to characterize the sensitivity of merger classifications to imaging quality.

\section{Results}
\label{Results:realism}

Having trained all five models and used them to make merger predictions on the test set galaxies, we now present the results, which have been separated into two main categories: same-survey results (Sections~\ref{Same-survey results} and~\ref{Performance trends with galaxy parameters}, in which models are evaluated on data with the same construction as their training sets) and cross-survey results (Section~\ref{Cross-survey results}, in which the potential for models to succeed outside of their training regime is explored). We will also pay particular attention to the trends of CNN performance with parameters that could influence the apparent strength of faint tidal morphologies, including redshift, depth, spatial resolution (both pixel scale and PSF), and merger mass ratio. In all cases, ``completeness'' refers to the fraction of galaxies belonging to a given class (usually the post-mergers) that are classified correctly by the model, while ``purity'' refers to the fraction of galaxies predicted by the model to be a given class that truly belong to that class.

\subsection{Same-survey results}
\label{Same-survey results}

The overall success of the CNNs in distinguishing post-mergers from controls can be summarized by conventional machine learning figures of merit for classification -- class-wise completeness (i.e., the fractions of post-mergers and controls that are classified correctly), receiver operating characteristic (ROC) curves, and purity-completeness curves. Each model is attempting to generalize over the intrinsic and observational diversity of the merger and control samples in order to correctly classify as many galaxies as possible. When models are less successful in classifying a certain subset of galaxies (e.g., one of the classes, or galaxies belonging to a certain mass range), it is because the model's overall performance was improved even as the subset in question was compromised.

\begin{figure*}
\includegraphics[width=\textwidth]{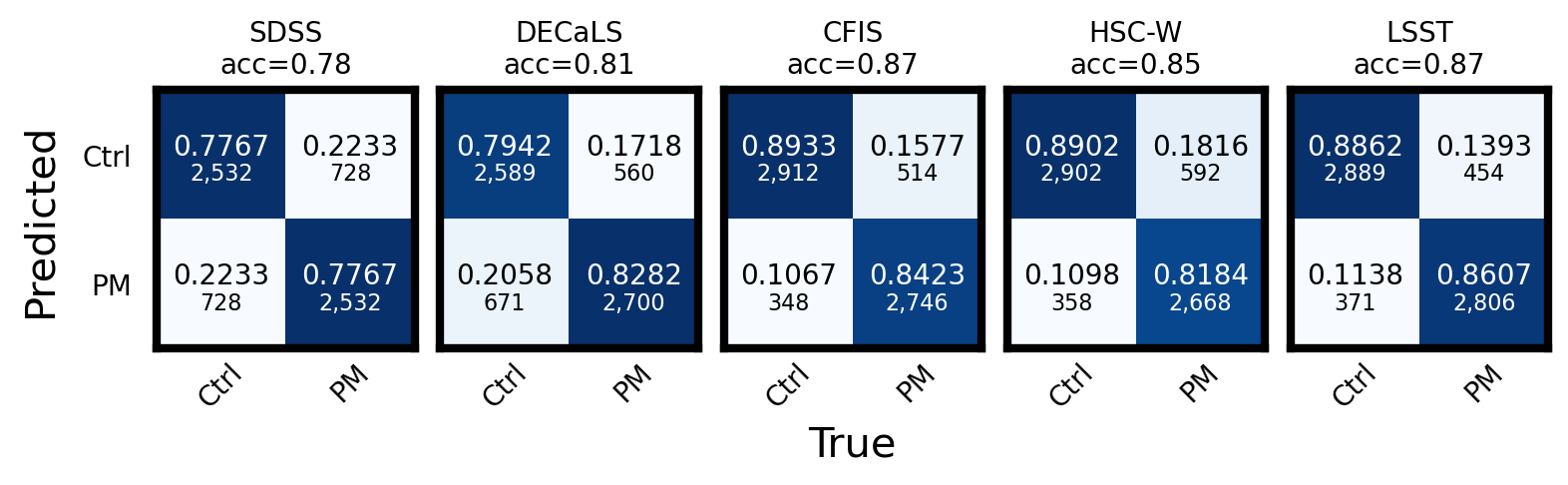}
\caption{Confusion matrices for the five trained CNN models, evaluated on test data with the same survey realism as their training data. The overall accuracy for each CNN (total fraction of galaxies labelled correctly) is shown above the panels. The completeness scores for post-mergers (bottom right corner of each confusion matrix) and on non-merger control galaxies (top left corner of each matrix) summarize the class-wise performance for each model.}
\label{fig:cmx_row_realism}
\end{figure*}

Figure~\ref{fig:cmx_row_realism} shows the confusion matrices for models trained and tested on data with realism based on the same survey. The surveys are arranged in order of increasing $5\sigma$ $r$-band limiting point-source depth from left to right and top to bottom. SDSS has the lowest total accuracy, as well as the lowest class-wise completenesses, for both post-mergers and controls. Based on the visual quality of the images, this is unsurprising -- galaxies at the highest $z$ studied cannot typically be separated by eye in SDSS (see Figures~\ref{fig:all_ims_merger} and~\ref{fig:all_ims_ctrl}), and visual separability is often a reasonable predictor of the potential of machine vision methods for a given problem. The exact balance between the post-merger and control classes in SDSS suggests that the visual degeneracy is equally confusing in both classes. In trying to adjust itself to account for the lack of merger features in many galaxies that are labeled as mergers, the model ultimately suffers a decrease in overall accuracy.

In DECaLS-quality imaging (representing the next step up from SDSS in terms of depth), the completeness is improved somewhat for both classes, but especially for the post-mergers, presumably since a larger number of galaxies with post-merger labels actually exhibit merger-like morphologies in the training data. Completeness statistics for both classes are again improved in CFIS imaging, since the morphological differences between mergers and controls are more visually distinct with better depth.

Referring to Figures~\ref{fig:all_ims_merger} and~\ref{fig:all_ims_ctrl}, one can begin to interpret the changes in performance for surveys with better $5\sigma$ limiting point-source depths than CFIS. At CFIS depth and better, the tidal tail feature in the merger image is visible across the entire redshift range, suggesting that CNNs will nominally be able to distinguish between the classes unless there is a spurious effect (e.g., a chance viewing angle from which a galaxy's tidal features are obstructed, or a merger that is intrinsically unusually faint). Even though the appearances of individual galaxies in HSC-W and LSST are distinct compared to CFIS, the relatively uniform visibility of merger-induced morphology across the three highest-quality datasets is likely responsible for the similarity in performance between the CNNs trained on CFIS, HSC-W, and LSST data. The specific role of depth and resolution in determining the efficacy of each CNN is investigated in greater detail later in Figures~\ref{fig:acc_v_5sig},~\ref{fig:acc_v_psf}, and~\ref{fig:acc_v_ccd}. In interpreting the global performance results, we emphasize that all five models have converged (without improving on the validation data set for 50 epochs).

There is also an apparent ceiling (at least for the degrees of survey realism studied here) for CNN performance for any class-wise completeness or global accuracy score at $\sim90$ per cent. The performance limit is consistent with the findings of \citet{2022MNRAS.511..100B}, who report completeness scores between $89-90$ per cent when idealized TNG100-1 stellar maps are used to train CNNs. \citet{2024arXiv240718396F}'s merger completeness would appear to be lower, at 84 per cent, but we note that the \citet{2024arXiv240718396F} merger class includes both pre- and post-merger galaxies in a wide temporal window. The comparison is therefore not exact. Since the galaxy morphologies are easily visible across the entire redshift range in the deepest data, we propose that this apparent limit is a consequence of intrinsic degeneracy between the stellar morphologies of galaxies belonging to the post-merger and control classes (e.g., some mergers have truly undisturbed appearances as a result of small $\mu$ or inconvenient viewing angles, regardless of image quality). Indeed, inspection of the five models' predictions for $\sim100$ individual galaxies reveals that the certain galaxies are intrinsically problematic, i.e., misclassified regardless of viewing angle, $z$, or survey due to intrinsic morphology. We do not compare the performance of the CNN on survey-realistic images to its performance on ``idealized'' data, since multiple $z$ realizations are not possible in a realism-free context. As such, the CNN would need to be trained on a significantly smaller dataset, and the comparison would no longer be direct. We can instead refer to \citet{2024MNRAS.528.5558W}, who include an idealized dataset in their experiment. Even though a different method is used, \citet{2024MNRAS.528.5558W} also find that excessive depth is not necessarily beneficial to merger identification. An analogous effect is therefore plausibly responsible for the results of this study.

\begin{figure*}
\includegraphics[width=\textwidth]{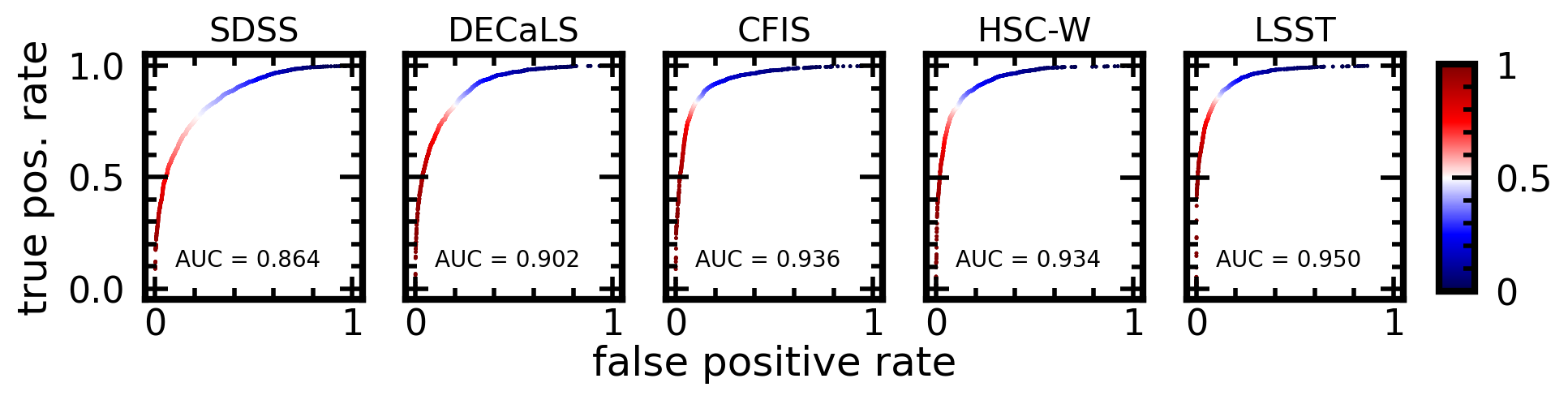}
\caption{ROC curves, another merit diagram for CNN classifiers, for the five CNNs. ROC curves plot the true positive rate (fraction of post-mergers correctly identified) and the false positive rate (fraction of incorrectly classified post-mergers) as a function of model decision threshold. The area under the ROC curve is also a figure of merit, with 0.5 equivalent to random performance, and 1.0 indicating perfect separation between the classes. ROC AUC score also increases generally with depth, reflecting again the importance of imaging depth in merger identification.}
\label{fig:roc_row_realism}
\end{figure*}

We next use ROC curves to collapse each model's performance down to a single figure of merit, an area under the curve (AUC) score. Figure~\ref{fig:roc_row_realism} shows ROC curves for the five CNNs, with the AUC score shown in each panel. The curves generally enclose more area with increasing depth, particularly in SDSS (with AUC=0.864), DECaLS (AUC=0.902), CFIS (AUC=0.936), and LSST (AUC=0.950) imaging. The AUC score for HSC-W (AUC=0.934) is slightly lower than for CFIS, possibly due to loss of some post-mergers at low-$z$ (see also Figure~\ref{fig:acc_v_z}).

\begin{figure*}
\includegraphics[width=\textwidth]{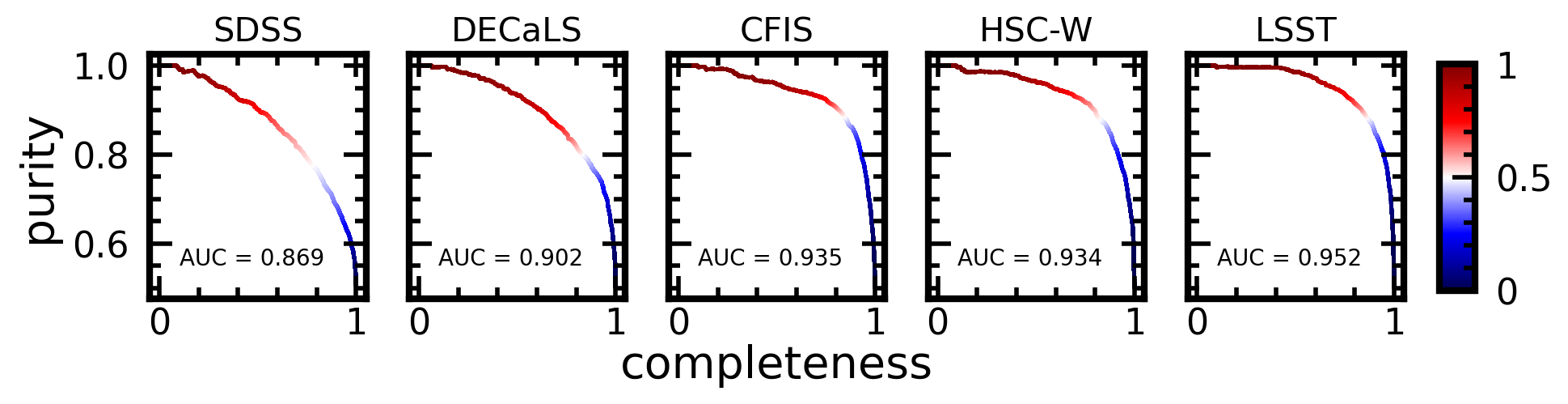}
\caption{Purity-completeness (or precision-recall, in machine learning parlance) curves for all five models evaluated on like data. Each panel shows the purity (or precision of the predicted merger sample) and completeness (recall of the predicted merger sample) as a function of the model's decision threshold (shown on the colour bar). The area under the curve is also a figure of merit, with an AUC of 0.5 indicating performance consistent with random, and an AUC of 1.0 indicating perfect separation between the classes by the model. The area under the curve scales generally with the depth of the survey studied, indicating that the ability of our CNN architecture to identify samples that are degrees of pure and complete depends on the observability of faint features in the images.}
\label{fig:prc_row_realism}
\end{figure*}

We also show the purity-completeness curves (PCCs) and the corresponding AUCs for all five CNNs in Figure~\ref{fig:prc_row_realism}. Purity-completeness curves are distinct from ROC curves in that they highlight the model's ability to return a sample of post-mergers that is pure and complete as a function of CNN $p(x)$. The CFIS-trained model (AUC=0.94) impressively achieves strong completeness for both classes at a shallower depth. The same-survey figures of merit (completeness scores on post-mergers and controls, and the areas under the ROC and purity-completeness curves) for all five models are summarized in Table~\ref{surv-merit-table}. Since the models are evaluated on equal-sized samples of post-mergers and controls, we note that the purity statistics reported in Table~\ref{surv-merit-table} would be lower if the models were used to evaluate a galaxy sample with realistic proportions of mergers and non-mergers.

\begin{table*}
\begin{center}
\begin{tabular}{ |c|c|c|c|c|c|c| } 
\hline
Survey & \ Post-merger completeness & \ Control completeness & \ Accuracy & \ Purity & \ Area under ROC & \ Area under PCC \\
\hline
\hline
SDSS & 0.78 & 0.78 & 0.78 & 0.78 & 0.86 & 0.87 \\
\hline
DECaLS & 0.83 & 0.79 & 0.81 & 0.80 & 0.90 & 0.90 \\
\hline
CFIS DR5 & 0.84 & 0.89 & 0.87 & 0.89 & 0.94 & 0.94 \\
\hline
HSC-W & 0.82 & 0.89 & 0.85 & 0.88 & 0.93 & 0.94 \\
\hline
LSST 10y & 0.86 & 0.88 & 0.87 & 0.88 & 0.95 & 0.95 \\
\hline
\hline
\end{tabular}
\end{center}
\caption{The class-wise completeness, overall accuracy, and post-merger purity statistics, as well as AUC scores for the ROC and purity-completeness curves for the CNNs trained for each survey. Figures are reported for equal-sized samples of mergers and non-mergers, so the purity statistics would be lower for all surveys if the CNNs were applied to a test set with a realistic distribution of merger stages.}
\label{surv-merit-table}
\end{table*}

\begin{figure}
\includegraphics[width=\columnwidth]{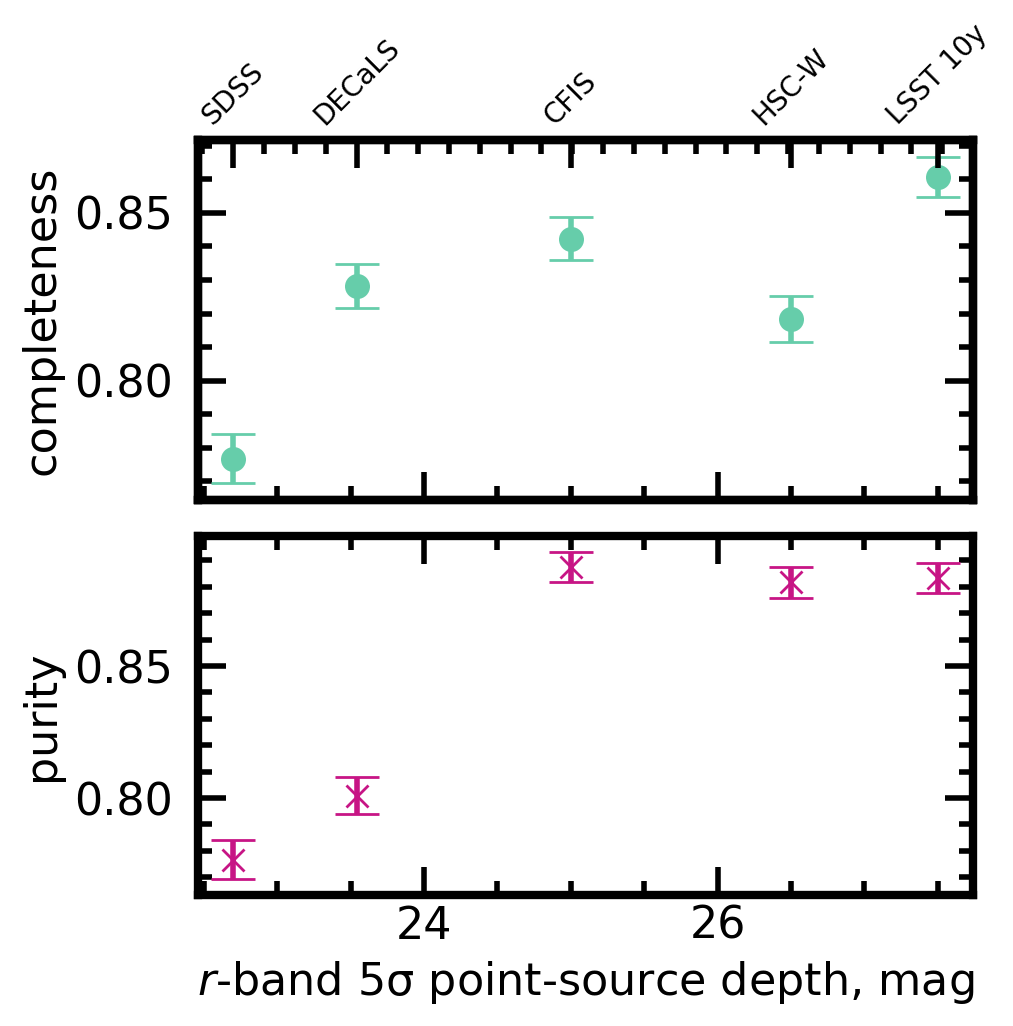}
\caption{The completeness and purity scores for models trained with five different survey realism parameters as a function of the reported $5\sigma$ limiting point-source depth for each survey. Completeness on the post-merger class is shown in teal, and the purity of the predicted post-merger sample is shown in magenta. Generally, deeper imaging is helpful to completeness and purity, but there is a diminishing return and additional complications in imaging from HSC-W, above the depth of CFIS.}
\label{fig:acc_v_5sig}
\end{figure}

Having stated the overall figures of merit, we will present the performance of the classifiers as a function of several quantities that we expect would bear on a CNN's ability to distinguish between mergers and controls. Figure~\ref{fig:acc_v_5sig} shows the post-merger completeness (teal data series) and the purity of the predicted post-merger sample (magenta data series) for the five CNNs, plotted as a function of the reported $r$-band limiting $5\sigma$ point-source depth. Viewing the performance metrics presented in Figure~\ref{fig:cmx_row_realism} in the context of each survey's limiting depth, it is clear that sufficiently deep imaging is helpful for reliable merger classifications, but that there is a diminishing return in taking observations deeper than $\sim25$ mag. As a result, the CNN is already finding mergers at low-$z$ in CFIS imaging at a success rate similar to what is expected for HSC-W and LSST.

\begin{figure}
\includegraphics[width=\columnwidth]{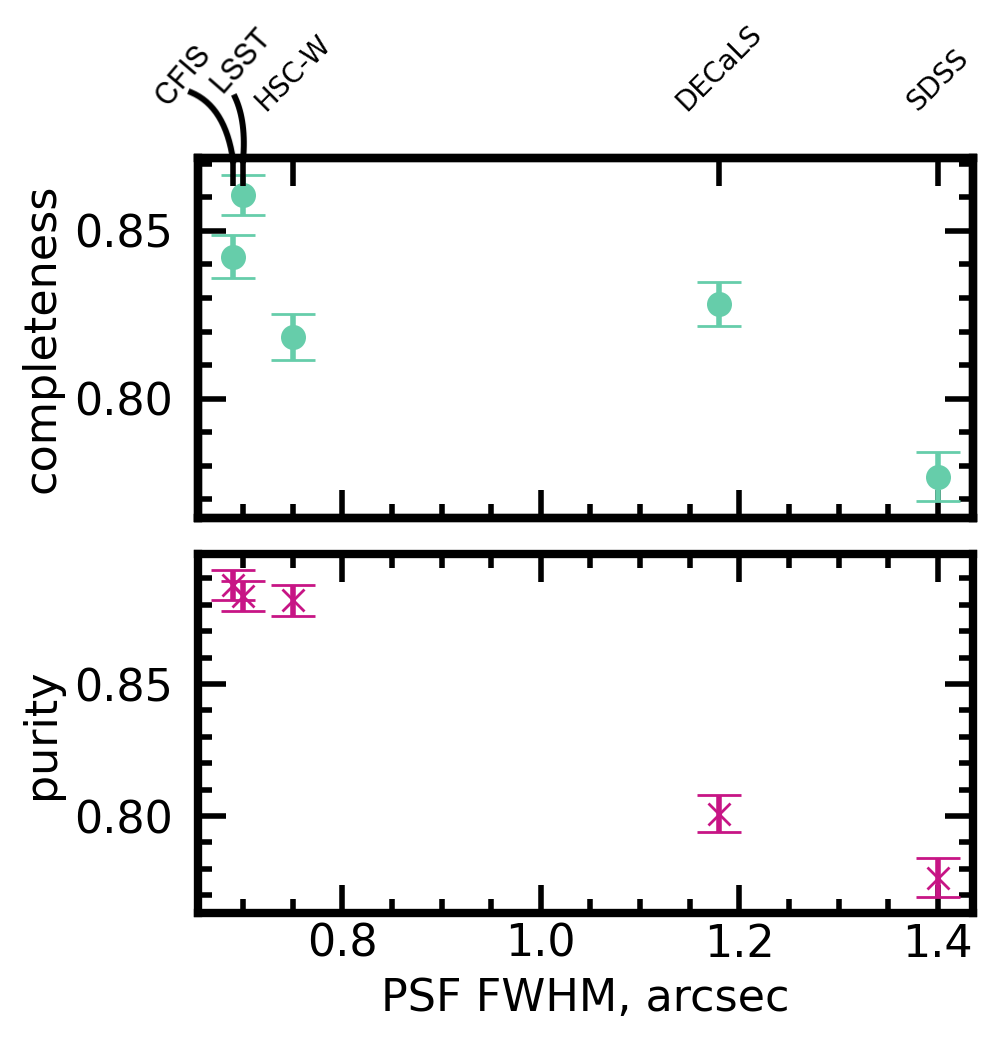}
\caption{The same as Figure~\ref{fig:acc_v_5sig} but for the typical PSF ($r$-band seeing) for each survey, which constrains the effective angular resolution. The completeness and purity achieved by the LSST model, HSC-W model, and CFIS model are closely grouped together on these axes, suggesting that reliable merger classifications can be completed for the $z$ range studied even when the PSF is atmosphere-dominated.}
\label{fig:acc_v_psf}
\end{figure}

Figure~\ref{fig:acc_v_psf} has the same construction as Figure~\ref{fig:acc_v_5sig}, but instead plots the surveys' performance metrics against the reported $r$-band seeing PSF in arcseconds. The PSFs for all of the surveys are dominated by contributions from Earth's atmosphere. There seems to be a trend between PSF and completeness, but we argue that this is mainly because survey depth generally trends with PSF (i.e., state of the art surveys are improved in both depth and resolution compared to legacy surveys; see Figure~\ref{fig:depth_res_grid} below). While it is difficult to estimate the specific contributions of depth and PSF to the final completeness of each model without generating a more extensive grid of mock models, we refer to \citet{2024MNRAS.528.5558W}, who find that depth is the main determinant of merger identification performance when combining non-parametric morphological parameters in a random forest classifier.

\begin{figure}
\includegraphics[width=\columnwidth]{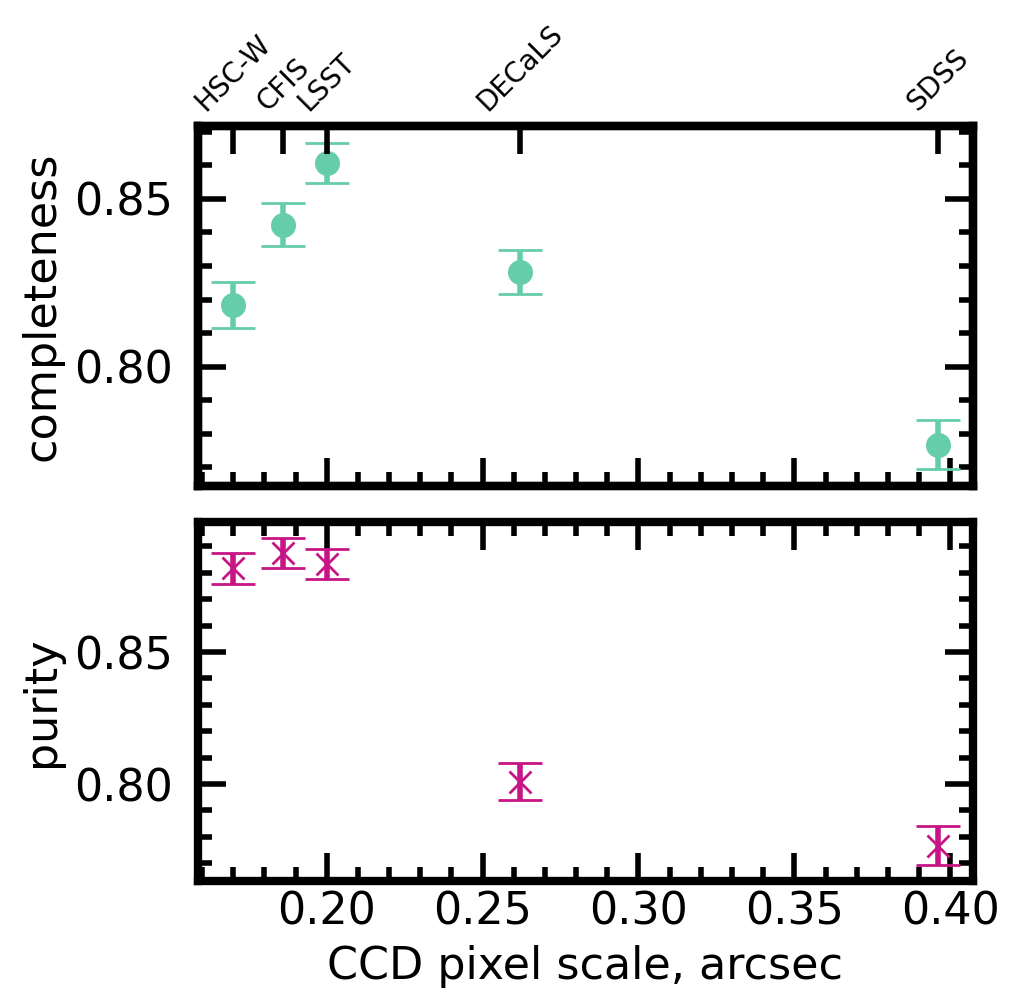}
\caption{The same as Figure~\ref{fig:acc_v_5sig} but for the angular pixel scale of the charged couple device (CCD) for each survey, which bears on both effective angular resolution and the pixel-wise S/N. The post-merger completeness and purity for the LSST model, HSC-W model, and CFIS model are grouped together on these axes as well.}
\label{fig:acc_v_ccd}
\end{figure}

The angular scale of the individual pixels of the camera CCD used to take images for each survey also bears on effective resolution and depth; Figure~\ref{fig:acc_v_ccd} plots the same performance metrics for the five models again but as a function of CCD pixel size. The trend with CCD scale is interesting for two reasons. First, the CCD scale determines the extent to which the maximum spatial resolution (set by the PSF) is preserved in the final image; in this way, it is a more direct determinant of spatial resolution than the PSF itself. Second, it plays a role in determining the effective S/N of the image, since signal and noise photons are effectively binned on a pixel-wise basis. Completeness increases with decreasing pixel scale from SDSS to DECaLS and from DECaLS to LSST, but turns down again for smaller pixel scales. It is unlikely that CCD pixel scale is the main driver of the final completeness and purity scores for each model, but they reflect the two main trends suggested by the results up to this point: that increasing depth is generally good for performance, but that an excess of spatial detail at the expense of pixel-wise S/N can be a source of confusion.

Since the morphological disturbances of significant (with $\mu>$1:10) mergers are generally on the scale of at least several kpc, we propose that imaging sensitivity (determined by the depth and CCD scale) plays a more important role in limiting performance than PSF, at least for the range of PSF studied here. The CCD pixel scale's role in determining the sensitivity of images is likely more important than its bearing on spatial resolution in the context of this study, since our image preparation pipeline involves resizing all images to $128\times128$ pixels. At low-$z$, the resizing operation also reduces the resolution of the images evaluated by the CNN (i.e., the resizing operation is the limiting factor on resolution), again de-prioritizing the resolution limit set by the CCD except in the highest-$z$ realization studied here. In other words, at a given survey depth, smaller CCD scale leads to lower pixel-wise S/N. It is this effect, rather than the spatial resolution constraint set by the CCD, that is leading to variable performance as a function of CCD scale.

\begin{figure}
\includegraphics[width=\columnwidth]{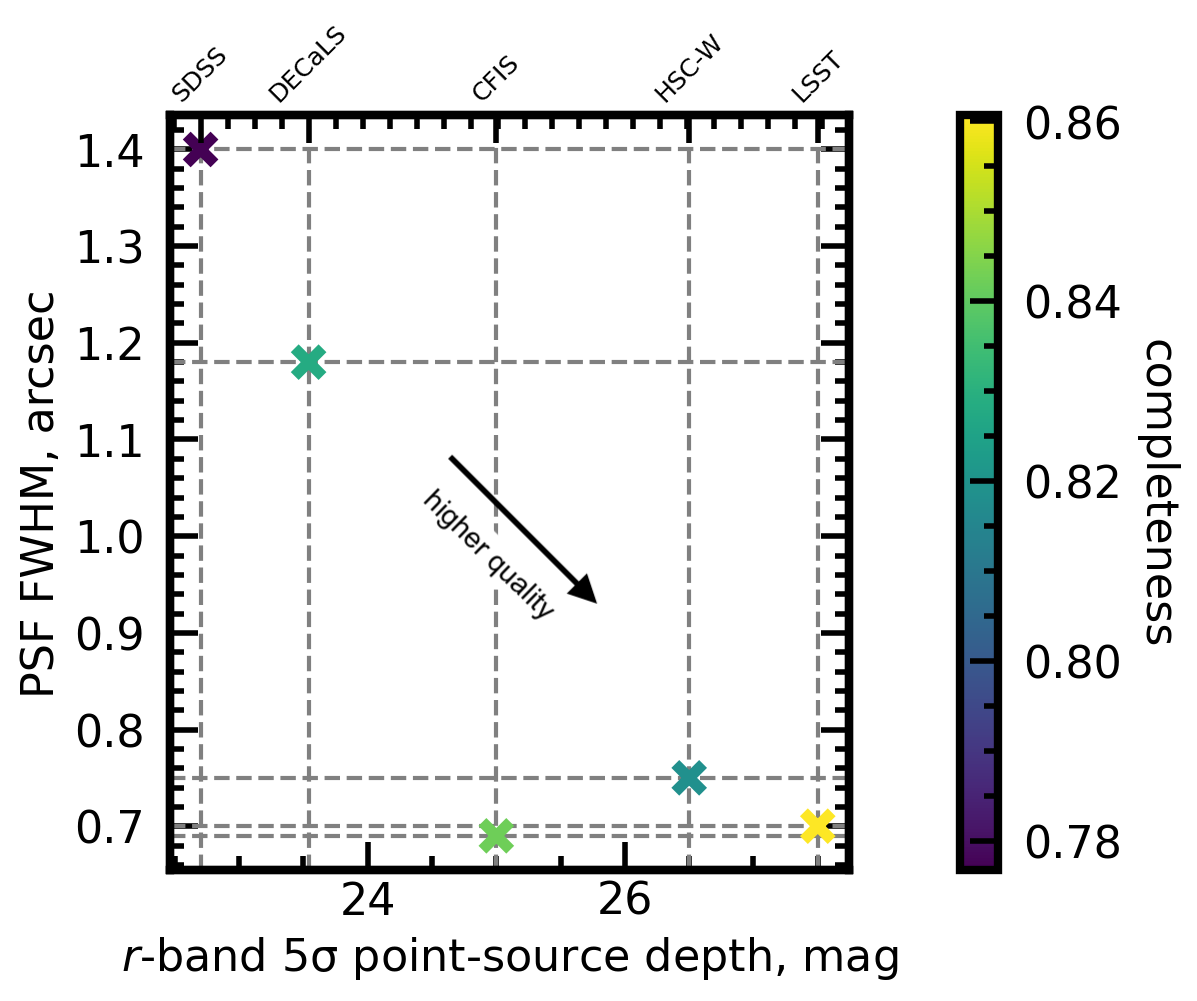}
\caption{The post-merger completeness scores for each of the five models (colour scale) plotted in the depth (here approximated by limiting $5\sigma$ point-source depth) and resolution (PSF FWHM in arcseconds) plane. Performance is sensitive to both parameters, with an apparent minimum limiting depth of $\sim24-25$ mag being important to high completeness. Resolution also plays a role in setting the final completeness score for each model, but the shallowest two surveys (SDSS and DECaLS) also have the worst spatial resolution, making it somewhat difficult to disentangle the individual contributions of each parameter.}
\label{fig:depth_res_grid}
\end{figure}

Finally, we plot the models' post-merger completeness scores on a two-dimensional plane of depth and PSF resolution in Figure~\ref{fig:depth_res_grid}. The post-merger completeness statistics for each model are represented by the blue-green colour scale for each marker on the plane. Viewed in two dimensions, the relative importances of the trends outlined in Figures~\ref{fig:acc_v_5sig} and~\ref{fig:acc_v_psf} can be summarized. Since the SDSS- and DECaLS-trained models have the largest PSF and shallowest depth, it is difficult to determine the precise roles of depth and resolution in setting the final completeness score for each model. Still, the visual characteristics of the post-mergers in SDSS and DECaLS data (especially at high-$z$, see Figure~\ref{fig:all_ims_merger}) suggest that the surveys' comparatively shallow depth may set a practical upper limit on the visibility of tidal features.

\subsection{Performance trends with galaxy parameters}
\label{Performance trends with galaxy parameters}

In Section~\ref{Same-survey results}, we demonstrated that the efficacy of CNN models for merger and non-merger classification is sensitive to survey depth and resolution, but it is also useful to investigate the representativeness of the merger samples that are identified by the CNNs in each image set.

\begin{figure*}
\includegraphics[width=\textwidth]{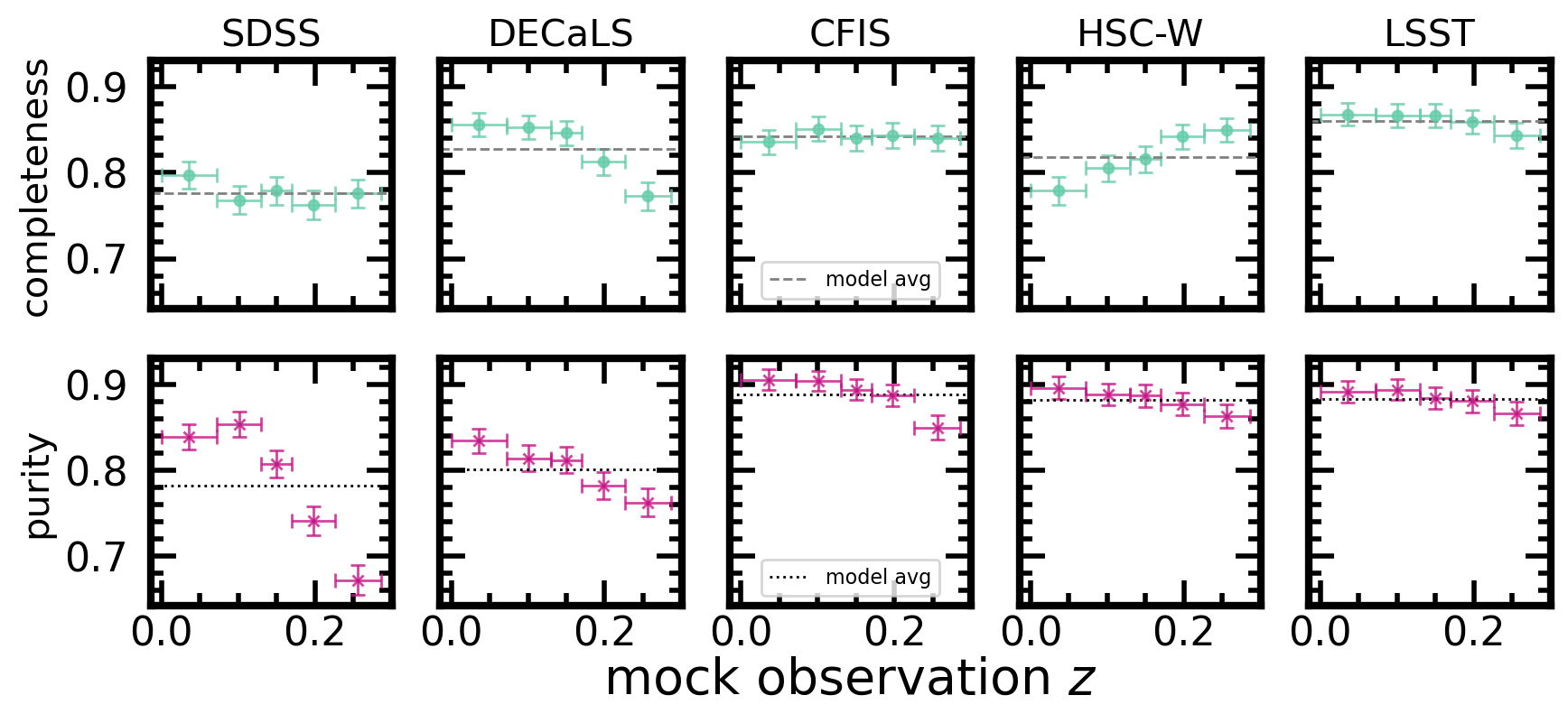}
\caption{The completeness of the five CNNs for post-mergers (teal), purity of the predicted post-merger samples (magenta), and the average completeness (grey dashed line) and average purity (black dotted line) for each model binned as a function of $z$. Statistics are reported at each of the five discrete redshifts used for mock observations (given in Section~\ref{Mock observations}). The error regions are the binomial errors on the statistics at each $z$. The diversity of the trends shown in each panel suggest that generalizing over the appearances of galaxies across a given $z$ range is difficult for CNNs.}
\label{fig:acc_v_z}
\end{figure*}

Low-surface brightness features fade with redshift as a result of cosmological dimming (as seen in Figure~\ref{fig:all_ims_merger}), so it is reasonable to expect that CNNs will struggle with calibration (learning to identify the range of surface brightnesses associated with tidal features) in shallow imaging and at high-$z$. The post-merger completeness and predicted post-merger sample purity for each of the five CNNs is shown as a function of mock observation $z$ in Figure~\ref{fig:acc_v_z}. The SDSS, CFIS, and LSST models retain approximately consistent completeness as a function of mock observation $z$. The DECaLS model identifies post-mergers more successfully at low-$z$, reflecting the limited visibility of merger-like morphology in the highest two redshift realizations in DECaLS imaging (see Figure~\ref{fig:all_ims_merger}). The HSC-W model has the opposite behaviour, possibly due to the resolution of the HSC CCD (see Figure~\ref{fig:acc_v_ccd}, which reveals decreasing performance with finer CCD resolution between LSST, CFIS, and HSC-W). We emphasize that decreasing merger completeness with CCD scale is not likely related to the limits on spatial resolution imposed by the CCD. Instead, the limit in pixel wise sensitivity at a given depth set by the CCD appears to be a more likely culprit.

Figure~\ref{fig:acc_v_z} also reveals that the purity of predicted post-merger samples decreases for all five CNNs with increasing $z$. This is to be expected, since at high-$z$, images in all five datasets contain larger random fluctuations due to sky noise. In some cases, we expect that these fluctuations and the muted appearance of some post-mergers' morphologies together give rise to more contaminated samples. The models' sensitivity to redshift is broadly the result of challenges inherent to the task of generalizing over a range of $z$.

\begin{figure*}
\includegraphics[width=\textwidth]{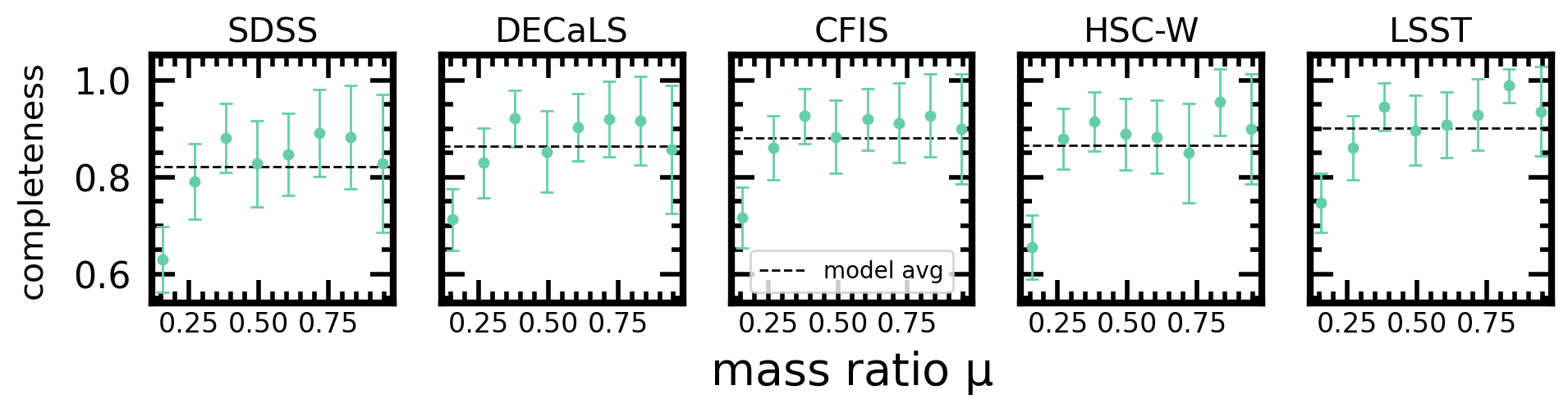}
\caption{The same as Figure~\ref{fig:acc_v_z}, but with purity and completeness statistics shown for galaxies arranged in eight bins of $\mu$. We find very similar trends as a function of $\mu$ across all five surveys, even though the models have shown themselves to behave very differently in other tracts of parameter space. The similarity of the trend in all five panels (increasing and stabilizing completeness for higher mass ratios) is intuitive, and suggests that merger mass ratio is one of the primary factors affecting whether a given galaxy will be selected as a post-merger. Post-mergers with $\mu<$1:4 are likely to be proportionally under-represented by some 20 per cent compared to the remnants of merger events with larger mass ratios. The data is only shown for post-mergers, since the mass ratios of long-past merger events for our control sample galaxies are not relevant.}
\label{fig:acc_v_mu}
\end{figure*}

The progenitor mass ratio $\mu$ is expected to be among the most important factors governing merger observability (see \citealp{2010MNRAS.404..575L}). Post-mergers whose progenitors were similar in mass are expected to be more dramatically disturbed compared to the remnants of more minor mergers (e.g., with $\mu<$1:10). Figure~\ref{fig:acc_v_mu} emphasizes this point, and suggests that the influence of $\mu$ on merger observability is the most ubiquitous one presented in this work. All five CNNs experience very similar trends (albeit with their amplitudes governed by the depth and resolution of each survey) with significant misclassification for post-mergers with $\mu<$1:4. Mergers near the lower limit of $\mu$ selection for the training set are therefore at risk of significant under-representation in the final merger sample predicted by the CNN, and the bias is likely worsened if visual classifications are used for quality control (e.g., as in \citealp{2021MNRAS.504..372B}) since more dramatic post-mergers are also more likely to be confirmed visually.

\begin{figure*}
\includegraphics[width=\textwidth]{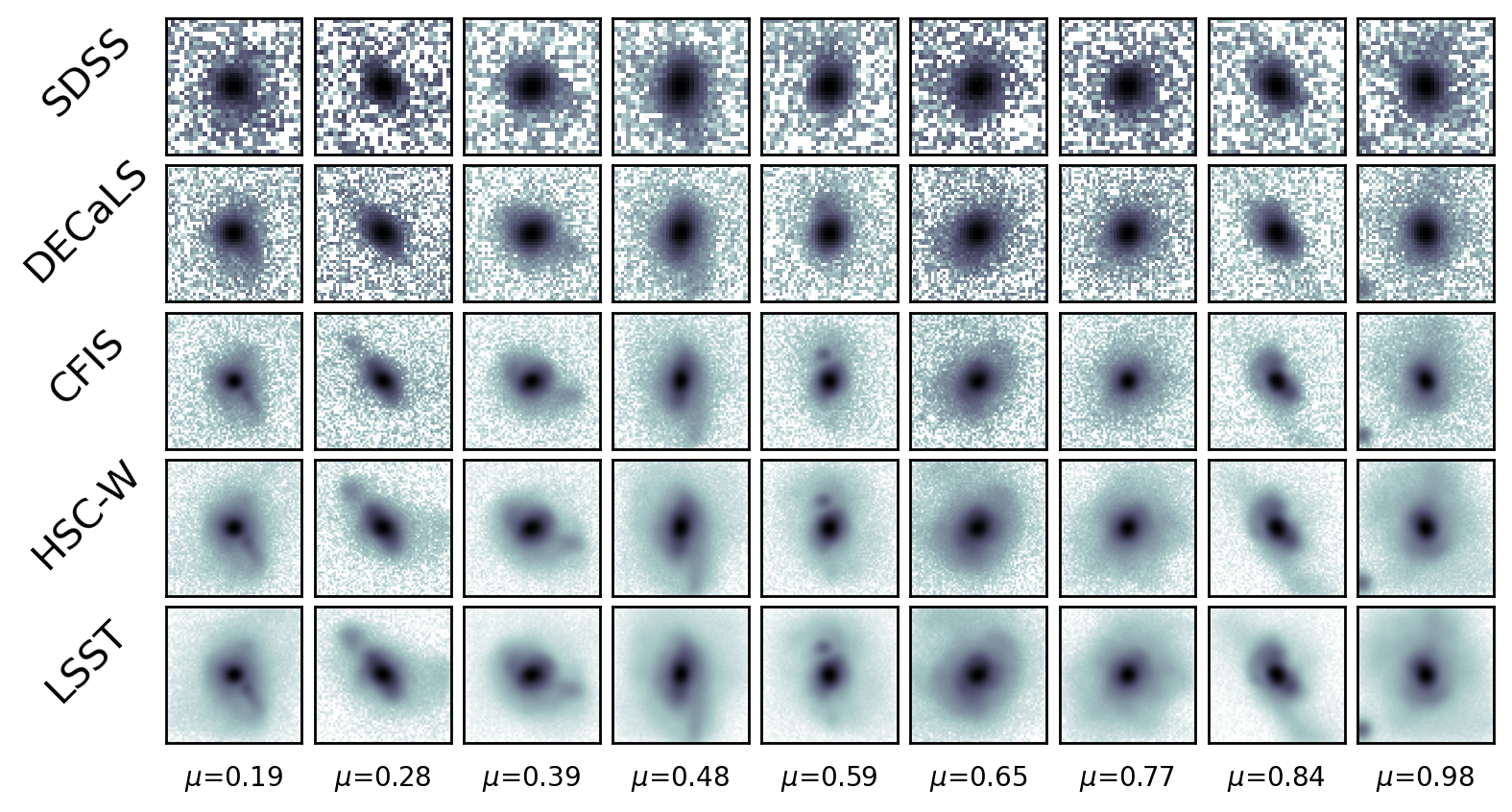}
\caption{In each row, we investigate the appearance of cases at $z=0.256$ (the highest $z$ realization studied here) for galaxies with increasing mass ratio $\mu$ from left to right. The $\mu$ values for the galaxies shown in each column are printed at the bottom of the figure. This figure demonstrates that the image quality associated with shallowest surveys (SDSS and DECaLS, especially) can obscure essential morphological information, potentially leading to misclassification.}
\label{fig:mu_highz}
\end{figure*}

The interplay between merger severity (as approximated by mass ratio, $\mu$) and $z$ is important to the results of this study, since we expect that small-$\mu$ mergers at high-$z$ are the most likely mergers to be overlooked by CNNs. We use Figure~\ref{fig:mu_highz} to investigate the relative importances of image quality and mass ratio at high-$z$. We select one galaxy with $10.4<\mathrm{log(M_{\star})/M_{\odot}}<10.6$ each from 9 equal-sized mass ratio bins between $0.1<\mu<1$, fix the redshift to $z=0.256$ (the highest redshift studied here), and visualize each galaxy in all five surveys. A narrow mass criterion is used to ensure relatively uniform total brightnesses in the galaxies being compared. Images for a given survey in Figure~\ref{fig:mu_highz} are shown in rows, while each column only contains images of one galaxy. Mass ratio increases from left to right, and survey depth increases from top to bottom.

Figure~\ref{fig:mu_highz} shows that merger-induced morphological disturbances are generally visible at $z=0.256$ in CFIS, HSC-W, and LSST, regardless of $\mu$ (at least for the cases shown). The visibility of merger-like morphology is a reasonable predictor of a CNN's ability to classify images correctly, but we emphasize that visibility does not translate a correct classification by the CNN in all cases. For DECaLS and SDSS, the two shallowest surveys included in the study, Figure~\ref{fig:mu_highz} suggests that the combination of small $\mu$ and high-$z$ may play a significant role in reducing the CNN performance statistics presented in Figure~\ref{fig:cmx_row_realism}.

\subsection{Cross-survey results}
\label{Cross-survey results}

The relative performance of the five CNNs presented in Section~\ref{Same-survey results} illustrate the potential of the ``best-case scenario'' in which models are trained and evaluated on data from the same image quality domain. But machine vision models (and deep learning models in general) cost time and substantial quantities of energy to train, both of which are valuable resources in science research (\citealp{2019arXiv190602243S}). The practice of transfer learning (i.e., training an already-trained model on a new dataset for fine tuning) shows promise in the galaxy classification domain (\citealp{2018MNRAS.479..415A,2019MNRAS.484...93D}), but transfer learning still mandates the creation of a new dataset and additional computing resources. If CNNs can be applied without re-training as a matter of course outside of their original domain without sacrificing scientific rigor, the efficiency of merger searches in the coming years could be improved substantially. Moreover, \citet{2024arXiv240117277B} already conducted a case study that indicated our CNN trained on IllustrisTNG galaxies with CFIS realism could be applied to shallower and lower-resolution imaging from DECaLS without incurring a prohibitive amount of loss or sample impurity. We will investigate the potential for merger searches using cross-survey classifications for the five trained CNNs described earlier in Section~\ref{CNN architecture and training strategy}.

\subsubsection{Overall out-of-domain performance}
\label{Overall out-of-domain performance}

\begin{figure*}
\includegraphics[width=\textwidth]{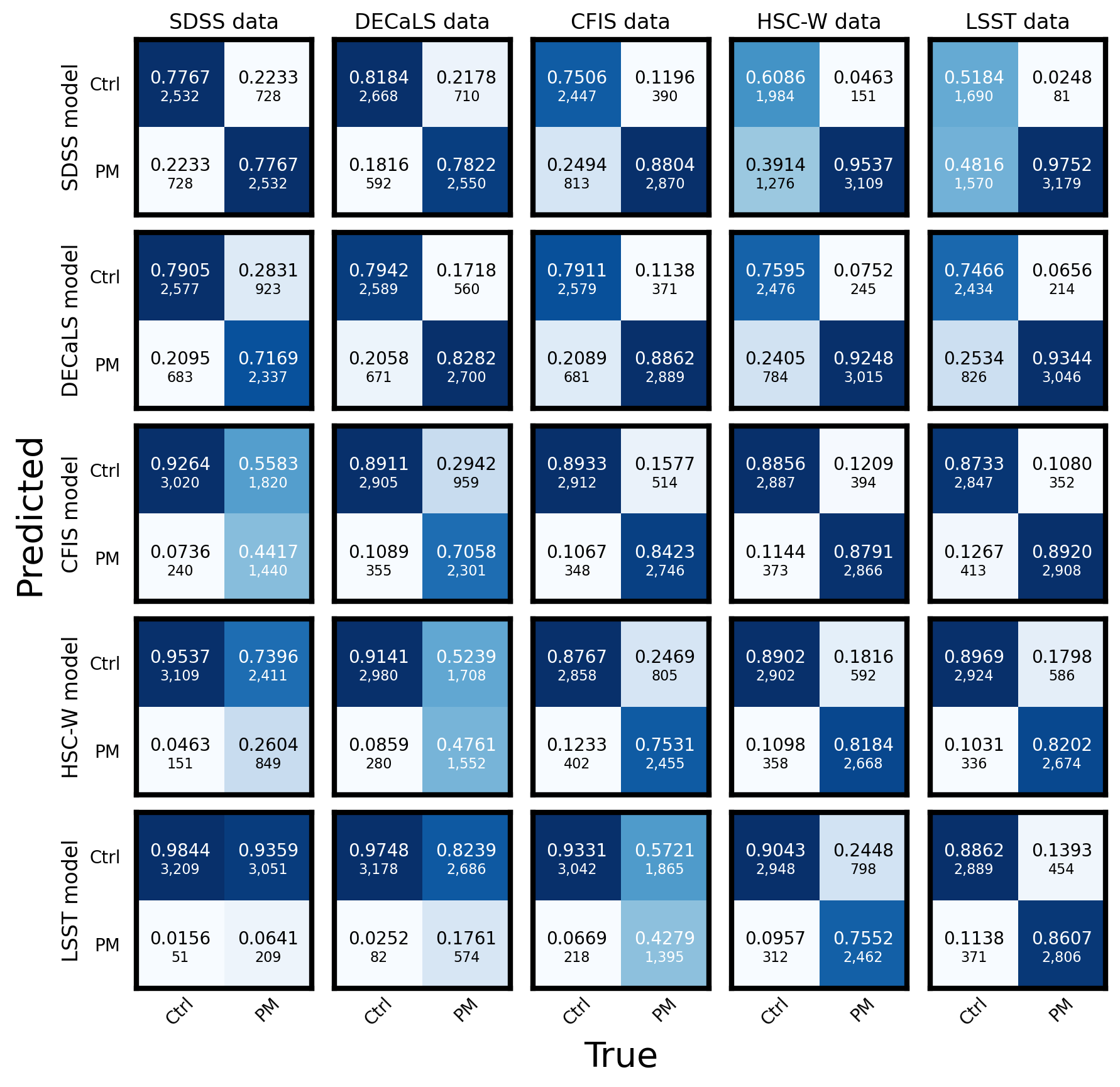}
\caption{Confusion matrices for all five trained CNNs after being applied to all five test datasets. Rows of matrices show the classification results for a single model (e.g., for the SDSS-trained model in the first row) on each of the test datasets, while columns show the classification results for each of the five trained CNNs on a given test set (e.g., for the CFIS test set in the third column). The confusion matrices on the diagonal are the same as shown in Figure~\ref{fig:cmx_row_realism}. Broadly, the matrices highlight the fact that calibration is essential for CNN-based merger searches. When models are applied to shallower data than their training set, they tend to under-predict mergers and over-predict controls. When models are applied to deeper data, they tend to misclassify a larger number of control galaxies as mergers.}
\label{fig:cmx_mx}
\end{figure*}

Figure~\ref{fig:cmx_mx} summarizes the results of the cross-survey inference experiment, in which all five CNNs are applied (without any re-training) to the test datasets for all five surveys. The classification results achieved by a single model are shown in rows, while the results for a given dataset are shown in columns. As in Figure~\ref{fig:cmx_row_realism}, the true labels are shown on the horizontal axes, and the machine-predicted labels are shown on the horizontal axes of each of the confusion matrices. The same five confusion matrices from Figure~\ref{fig:cmx_row_realism} are shown along the diagonal as well, for reference.

The results of the cross-survey inference experiment can be broadly characterized as an illustration of the importance of calibration for automated merger searches. If a model learns to identify the features of a recent merger event in a certain range of brightness, it will (perhaps intuitively) misclassify control galaxies if non-mergers as imaged by a different survey appear to have diffuse or asymmetric features with similar brightness levels. Conversely, if mergers in a different survey appear to lack tidal features within the learned range of brightness, a model will misclassify those mergers as controls. The importance of calibration is verified in the class-wise completenesses shown in the confusion matrices above and below the diagonal in Figure~\ref{fig:cmx_mx}. Above the diagonal, models are used to classify deeper imaging than their training regime. As a result, models typically mis-classify a large number of controls as post-mergers; for an extreme case, refer to the top-right confusion matrix, in which the SDSS-trained model misclassifies nearly half of the controls as post-mergers when applied to the LSST-like imaging. Below the diagonal, models are applied to shallower imaging than in training. As a result, a large number of post-mergers are mistakenly labeled by the models as controls; see for example the bottom-left confusion matrix, in which the LSST-trained CNN classifies the vast majority of galaxies as non-mergers regardless of their true label when applied to SDSS imaging. For higher-$z$ images in the DECaLS and SDSS datasets, the models trained on deeper imaging suffer two distinct penalties: the first imposed by the significant loss of merger features below the noise (see again Figures~\ref{fig:all_ims_merger} and~\ref{fig:mu_highz}), and the second imposed by mis-calibration. Calibration is an obstacle for all of the out-of-domain classification results presented here, and can (in theory) be remedied by re-training. The loss of visibility for mergers in shallow data is more difficult to overcome, and decreases the maximum potential of any model applied to data from shallow imaging surveys.

Figure~\ref{fig:cmx_mx} also reveals some surprises in the cross-survey inference experiment that depart from the general trend just described. The HSC-W model and CFIS model perform very well when applied to deeper domains than their training data, perhaps thanks to the plateau in the trend of performance versus $5\sigma$ limiting point-source depth shown in Figure~\ref{fig:acc_v_5sig}. Since the practical visibility of merger features (see again Figures~\ref{fig:all_ims_merger} and~\ref{fig:mu_highz}) do not appear to change significantly at depths better than $\sim25$ mag, we posit that the appearance of merger features become standardized after images are resized and normalized via the method described in Section~\ref{CNN architecture and training strategy}. In the context of a hybrid CNN and visual inspection merger identification framework, the CFIS and HSC-W models are therefore extremely useful, i.e., allowing for the recovery of a large proportion of post-mergers outside of their training regimes. In all cases, using visual inspection or other ensemble classification methods as a form of quality control \textit{a posteriori} is advisable.

The particular combination of training and inference data used in \citet{2024arXiv240117277B} is shown in the third row, second column, where the CFIS-trained model is applied to DECaLS data. The model misclassifies an additional 14 per cent of post-mergers compared to its performance in CFIS data, a result consistent with the lower global visual agreement fractions for post-mergers reported in \citet{2024arXiv240117277B} compared to \citet{2022MNRAS.514.3294B} for post-merger searches in CFIS and DECaLS, respectively. Still, the apparent purity of the post-merger sample from DECaLS and the results in Figure~\ref{fig:cmx_mx} indicate that merger searches can be conducted responsibly using cross-survey inference as long as additional quality control is enforced and any physical biases in the sample are accounted for (or at least discussed) in post.


\subsubsection{Changes in purity and completeness}
\label{Changes in purity and completeness}

\begin{figure}
\includegraphics[width=\columnwidth]{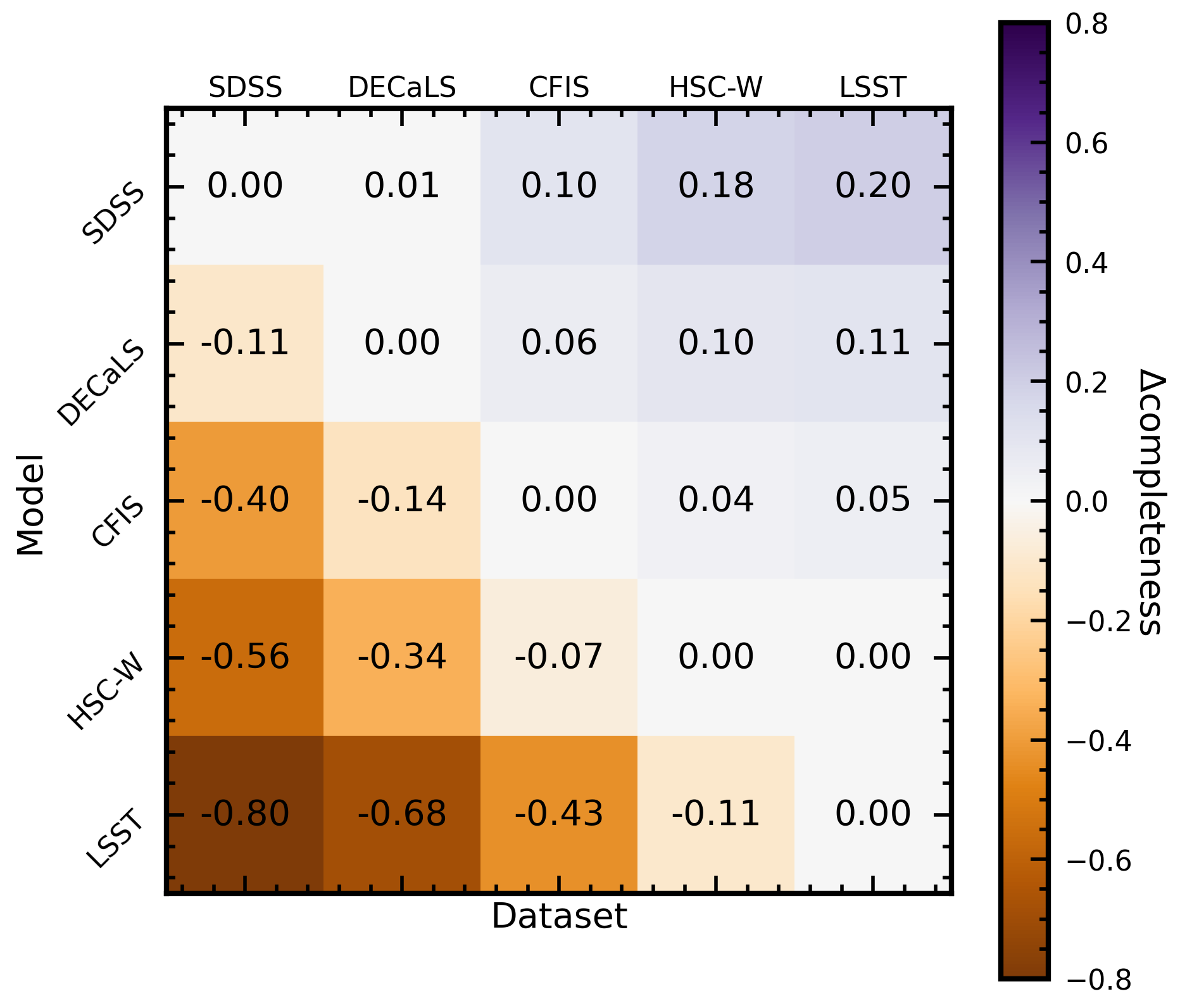}
\caption{The same configuration as Figure~\ref{fig:cmx_mx}, but reporting the change in post-merger completeness for each model when it is applied outside of its training domain compared to its same-survey completeness score. Positive $\Delta$ completeness (shown in purple) indicate that a model identifies a greater proportion of the true post-mergers in the test set, while negative $\Delta$ completeness (shown in ochre) indicate that the model identifies fewer post-mergers compared to its baseline. Values along the diagonal are zero by definition.}
\label{fig:delcomp_grid}
\end{figure}

Figure~\ref{fig:delcomp_grid} directly addresses one of the central questions of the cross-survey inference experiment: how much post-merger completeness is lost when a CNN is applied outside of its training domain? In each cell of Figure~\ref{fig:delcomp_grid}, we subtract the model's completeness score on a given dataset from its baseline completeness, i.e., its completeness score when applied to the test set from its training domain. Values along the diagonal are therefore zero by definition. Figure~\ref{fig:delcomp_grid} illustrates that completeness generally increases when models are applied to deeper imaging than their training domain (note the trend of positive $\Delta$ completeness above the diagonal, shown in purple) and decreases when they are applied to shallower imaging than their training domain (negative $\Delta$ completeness below the diagonal, ochre).



\begin{figure}
\includegraphics[width=\columnwidth]{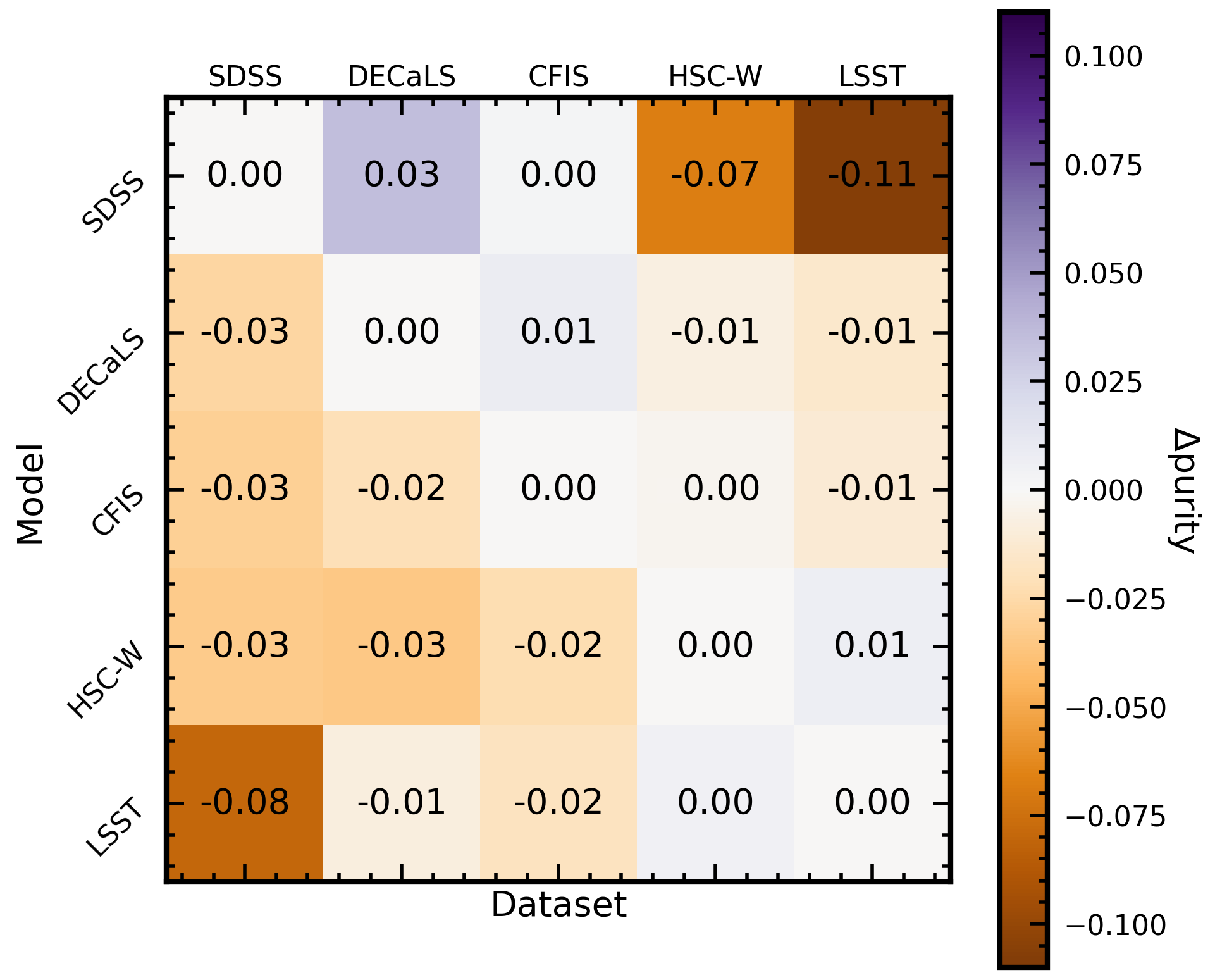}
\caption{The same as Figure~\ref{fig:delcomp_grid}, but reporting the change in the purity of the predicted post-merger sample for each model when it is applied outside of its training domain compared to its same-survey purity score. Values along the diagonal are zero by definition.}
\label{fig:delpur_grid}
\end{figure}

Figure~\ref{fig:delpur_grid} has the same construction as Figure~\ref{fig:delcomp_grid}, but shows the change in purity of the predicted post-merger sample when each model is applied outside of its training domain. The baseline purities for the same-survey experiment are given above in Table~\ref{surv-merit-table}. Figure~\ref{fig:delpur_grid} demonstrates that $\Delta$ purity is usually smaller than $\Delta$ completeness, though we note that this may be somewhat misleading: since we have created test datasets for each survey that include equal numbers of post-mergers and non-post-mergers, the purity scores presented throughout this work are artificially high compared to what could be achieved with a single CNN in observations.

Figure~\ref{fig:delpur_grid} indicates that purity is generally lost when models are applied outside of their training regime, but there are noteworthy exceptions to the rule. The SDSS-trained model improves in both purity and completeness (see Figure~\ref{fig:delcomp_grid}) when it is applied to DECaLS imaging. This is somewhat intuitive, since mergers in the DECaLS dataset at higher-$z$ are more likely to retain their merger-like appearance. In general, the results of the cross-survey inference experiment confirm that SDSS imaging is too shallow to conduct a holistic merger search spanning the survey's redshift domain.

\begin{figure*}
\includegraphics[width=\textwidth]{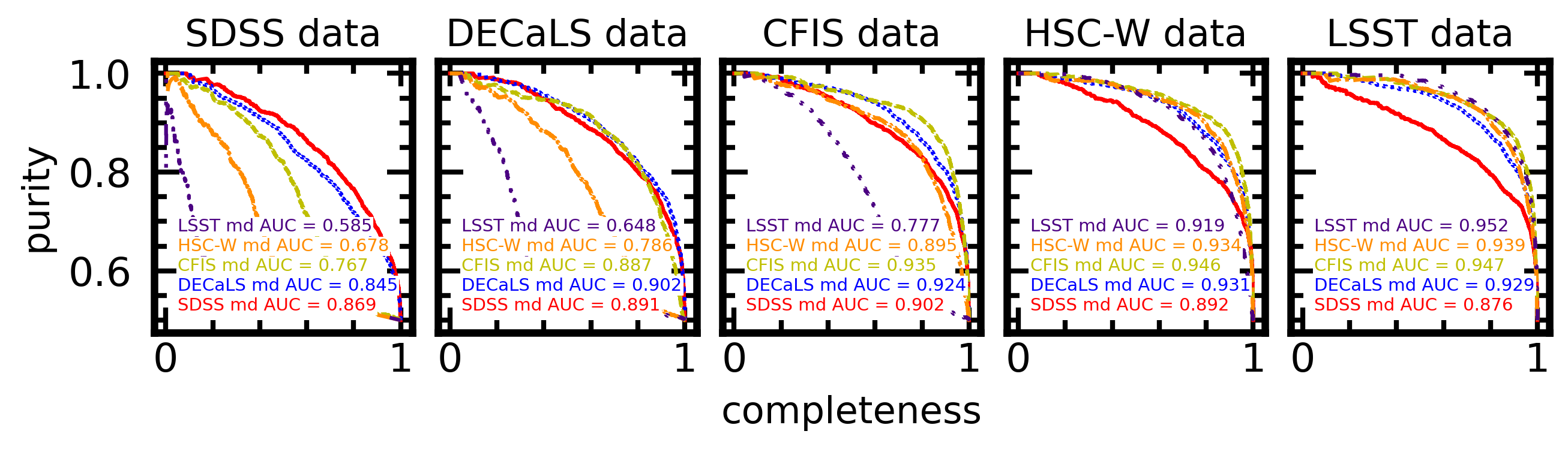}
\caption{Purity-completeness (or precision-recall) curves for the cross-survey inference experiment. The series of colour-coded cuves in each panel show the purity-completeness curves for each of the five models applied to one of the datasets. The annotations show the AUC scores for each curve, illustrating each model's potential to identify samples of post-mergers that are pure and complete when applied outside of their training regimes.}
\label{fig:prc_matrix}
\end{figure*}

Figure~\ref{fig:prc_matrix} shows the results of the cross-survey inference experiment in another way, folding in the utility of a cut in CNN $p(x)$ for each model. The purity-completeness curves are shown only for the post-merger classes, as the primary class of interest in the context of a merger search. Mirroring the results in Figure~\ref{fig:cmx_mx}, the curves for models applied to shallower data are typically lacking in completeness for a given $p(x)$, even though the samples they identify may be very pure. Improving overall performance by each of the models in deeper data again suggests the conclusion that there is an effective standardization of the appearance of merger features in deep imaging, allowing for more successful cross-survey inference. Still, the purity completeness curves indicate that models trained specifically to identify post-mergers in a given image survey are optimal in most cases.

\section{Discussion and conclusions}
\label{Discussion and conclusions:realism}

We have detailed the limits placed on low-$z$ merger searches by image quality, using five real surveys as benchmarks, and holding all other methodological variables constant. We find the relationship between image quality and the efficacy of CNNs to be complex. Intuitively, images must be deep enough to detect faint tidal features across the entire redshift range studied in order to be successful. Perhaps unexpectedly, greater depth is not categorically beneficial: in detecting recent (within one IllustrisTNG 100-1 simulation snapshot, some 150 Myr) and impactful (with $\mu>$1:10) mergers, we find that the depth of CFIS imaging (with a limiting 5$\sigma$ point-source depth of $\sim$25 mag in the $r$-band) is adequate. In other words, imaging with the depth and angular resolution of surveys like CFIS or HSC-W are already near the apparent limit of what is possible for binary classification for $\mu>$1:10 merger remnants (see also \citealp{2022MNRAS.511..100B}). Additional imaging depth is met with a diminishing return in completeness and purity. Additional detail detected in extremely deep (e.g., LSST after 10 years of co-adds) imaging may be beneficial for more advanced merger characterization tasks that are only lately being explored -- e.g., temporal estimation of merger timescales (see \citealp{2024A&A...687A..45P,2024arXiv240718396F}), identification of merger remnants with smaller progenitor $\mu$ values (i.e., ``mini mergers'' as studied in \citealp{2024MNRAS.527.6506B}), or characterization of fainter galaxies with lower stellar masses.

The results of this work are particularly useful when considered alongside the results of \citet{2024MNRAS.528.5558W}, who use non-parametric morphological statistics to perform a systematic analysis of merger recovery as a function of depth, resolution, viewing angle, and a variety of galaxy properties. \citet{2024MNRAS.528.5558W} used different observational realism and merger identification techniques than this work, but the trends of performance with depth and resolution still offer a useful point of comparison. We particularly refer the reader to Figure 10 of \citet{2024MNRAS.528.5558W}, which shows the result of an experiment very much like the one conducted in Section~\ref{Same-survey results}. Wilkinson et al. incorporate multiple non-parametric morphological statistics (asymmetry, outer asymmetry, shape asymmetry, Gini, and M20) in a random forest classifier, which is used to predict merger status. The approach of \citet{2024MNRAS.528.5558W} is quite similar to the one used in this work, in that CNNs also combine a feature extraction component (although the features in the CNN are learned in the convolution layers, rather than prescribed) with a machine learning classification tool (the fully connected layer of the CNN).


\citet{2024MNRAS.528.5558W} identify two main trends that are echoed in this work. First, greater depth at a given resolution is generally beneficial, but that the improvement in completeness and purity with increasing 5$\sigma$ limiting point-source depth above 25 mag is marginal compared to the improvement between $23-25$ mag.  \citet{2024MNRAS.528.5558W} also report a turnover in performance as a function of PSF FWHM, with the peak completeness recovered in imaging with a PSF of 0.75 arcsec. The other results on the depth-resolution plane explored in \citet{2024MNRAS.528.5558W} offer helpful context by filling in the bigger picture for our results, since the five surveys studied here represent discrete selections from a multi-variate grid of depth, resolution, and CCD scale.

While CCD scale is held constant in \citet{2024MNRAS.528.5558W}, the parameter may bear significantly on the results presented here, since the CCD scale sets the value of S/N in each pixel for a given observation by binning both the signal and the noise. Even though the final $128\times128$ image resolution used to prepare images for the CNNs generally dominates over the CCD scale, the resizing operation does not bear on spatial sensitivity in the same way that the CCD scale does. We use the \textsc{skimage} resize algorithm, which stretches or shrinks images, rather than re-binning them, to prepare images for training, so there is no impact on spatial sensitivity. In practice, a larger choice of CCD scale could enable the detection of a faint tidal feature at the expense of spatial detail, while a smaller CCD scale could make the opposite exchange. Indeed, this effect may be responsible for the decreased completeness exhibited by the CNN trained on HSC-W images in Figure~\ref{fig:acc_v_ccd}. More broadly, we posit that finer pixel-scale resolution may result in loss of post-merger completeness for surveys with sufficiently deep 5$\sigma$ limiting point source depths. For astronomers interested in searching for $\mu>$1:10 merger remnants below $z<0.3$ using deep imaging and a fine CCD, our results indicate that re-binning to a pixel scale of $\sim0.2$ arcsec will be beneficial. Re-binning will introduce a helpful amount of morphological smoothing, and the spatial sensitivity of the final image will also be improved.

We note that the results above lack an assessment of Euclid, the forthcoming space-based imaging survey. Euclid will also have ample depth (a limiting 5$\sigma$ point source depth of 26.2 mag in the visible band) and an excess of spatial resolution compared to the surveys included in our main results (PSF of 0.18 arcsec, CCD scale of 0.1 arcsec; see \citealp{2022A&A...662A.112E}). The results for Euclid are not shown since the softening length of IllustrisTNG100-1 (at $\sim1$kpc) dominates over the PSF of Euclid across the entire redshift range except for the realization at $z$=0.256. Including results from the Euclid experiment would therefore be distinct from the rest of our results (for which the PSF dominates the effective resolution), and therefore could be misleading. Euclid could be evaluated in the future, however, using a higher redshift analogue to the study conducted here, or using a higher-resolution simulation suite for training.


The figures of merit for the CNN trained to identify tidal features in HSC images in \citet{2023MNRAS.521.3861D} (overall completeness of 0.84, merger purity of 0.72, and merger class completeness of 0.85) are in reasonable agreement with the results for HSC in this work (overall completeness of 0.86, merger purity of 0.82, merger completeness of 0.82), though the HSC-W model presented here is able to improve on the purity of the predicted merger class via a statistical ``sacrifice'' of 3 per cent completeness on the merger class and an overall completeness that is higher by 2 per cent. \citet{2023MNRAS.521.3861D} also report significant loss when their model is applied to real galaxies imaged in HSC. While this work offers no direct comparison regarding the performance of simulation-trained models to HSC imaging, visual classification exercises have revealed (e.g., in \citealp{2022MNRAS.514.3294B}) that the CFIS-trained CNN used to identify the post-mergers performs well (qualitatively) in the observational domain. \citet{2023MNRAS.521.3861D} suggest that their loss of performance when classifying real galaxies may be the result of the fact that contaminating sources and image artefacts (e.g., foreground stars, zero-flux artefacts) are not included in the training set (see also \citealp{2019MNRAS.490.5390B}). It is therefore likely that the proof-of-concept models trained for this work would not perform well in the observational domain either -- since the results presented here are mainly intended as a methodological comparison, no artefacts were included in the training data used.

The results presented in \citet{2022MNRAS.513.1459M} indicate that 10-year depth LSST imaging taken at the Vera Rubin observatory should capture some $60-80$ per cent of the flux from tidal features in Milky Way-mass (or greater mass) galaxies at $z\sim0.05$, depending on the assumed final depth of the survey. To the extent that simulated galaxies processed with LSST observational realism have similar appearances to real galaxies in LSST, one would expect LSST to be an excellent opportunity for merger searches using a simulation-based approach. While LSST may be able to capture a groundbreaking amount of information about galaxies' recent assembly histories in the form of low-surface brightness detail, the results presented in this work indicate that LSST-quality imaging is not required for the identification of a pure and complete sample of recent mergers. At high-$z$, however, the potential of LSST for merger identification is substantially greater -- with more frequent merger events at earlier times in the Universe, the boundaries between mergers and non-mergers will be unusually blurred. LSST can therefore be used to collect the larger and higher-quality samples required to marginalize over merger status in the way that is currently possible at low-z with CFIS. We find that the change in the CNN's performance between CFIS and LSST is marginal, within $\sim2$ per cent for both mergers and non-mergers. We therefore argue that there is little benefit in waiting for 10 years of co-adds from LSST -- our results indicate that large-scale merger searches will be effective in the first few years of the survey.

In light of the fact that ground-based surveys like CFIS and HSC are already available, we posit that images from next-generation surveys like LSST and Euclid can be used to investigate more difficult and granular questions in galaxy evolution astronomy; for example, probing the impact of mergers with smaller mass ratios, at higher redshift, and / or with greater temporal specificity. Space for improvement remains for $\mu>$1:10 mergers at low-$z$ (i.e., the domain studied in this work), but the potential of next-generation hydrodynamical simulations and imaging surveys for temporal, high-$z$, and small $\mu$ merger characterization is much greater.

\section*{Acknowledgements}
\label{Acknowledgements}

The work detailed above was conducted at the University of Victoria. We acknowledge with respect the Lekwungen peoples on whose unceded traditional territory the university stands, and the Songhees, Esquimalt and WSÁNEĆ peoples whose historical relationships with the land continue to this day.

CFIS is conducted at the Canada-France-Hawaii Telescope on Maunakea in Hawaii. We also recognize and acknowledge with respect the cultural importance of the summit of Maunakea to a broad cross section of the Native Hawaiian community.

CB gratefully acknowledges support from the Forrest Research Foundation.

This work is based on data obtained as part of the Canada-France Imaging Survey, a CFHT large program of the National Research Council of Canada and the French Centre National de la Recherche Scientifique, and on observations obtained with MegaPrime/MegaCam, a joint project of CFHT and CEA Saclay, at the Canada-France-Hawaii Telescope (CFHT) which is operated by the National Research Council (NRC) of Canada, the Institut National des Science de l’Univers (INSU) of the Centre National de la Recherche Scientifique (CNRS) of France, and the University of Hawaii. This research used the facilities of the Canadian Astronomy Data Centre operated by the National Research Council of Canada with the support of the Canadian Space Agency.

Data from the IllustrisTNG simulations are integral to this work. We thank the Illustris Collaboration for making these data available to the public.

Funding for the SDSS and SDSS-II has been provided by the Alfred P. Sloan Foundation, the Participating Institutions, the National Science Foundation, the U.S. Department of Energy, the National Aeronautics and Space Administration, the Japanese Monbukagakusho, the Max Planck Society, and the Higher Education Funding Council for England. The SDSS Web Site is http://www.sdss.org/. The SDSS is managed by the Astrophysical Research Consortium for the Participating Institutions. The Participating Institutions are the American Museum of Natural History, Astrophysical Institute Potsdam, University of Basel, University of Cambridge, Case Western Reserve University, University of Chicago, Drexel University, Fermilab, the Institute for Advanced Study, the Japan Participation Group, Johns Hopkins University, the Joint Institute for Nuclear Astrophysics, the Kavli Institute for Particle Astrophysics and Cosmology, the Korean Scientist Group, the Chinese Academy of Sciences (LAMOST), Los Alamos National Laboratory, the Max-Planck-Institute for Astronomy (MPIA), the Max-Planck-Institute for Astrophysics (MPA), New Mexico State University, Ohio State University, University of Pittsburgh, University of Portsmouth, Princeton University, the United States Naval Observatory, and the University of Washington.

This research was enabled, in part, by the computing resources provided by Compute Canada.

\section*{Data Availability}
\label{Data Availability}

Simulation data from TNG100-1 used in the generation of training images for this work are openly available on the IllustrisTNG website, at tng-project.org/data. Template versions of \textsc{RealSim} and \textsc{RealSim-CFIS}, developed by Connor Bottrell with modifications by RWB are publicly available via GitHub at github.com/cbottrell/RealSim and github.com/cbottrell/RealSim-CFIS. Specific image training data used to develop the findings of this study are available by request from RWB.



\bibliographystyle{mnras}
\bibliography{cnn-ill-pm} 



\bsp	
\label{lastpage}
\end{document}